\documentclass[journal,comsoc]{IEEEtran}

\usepackage[T1]{fontenc}% optional T1 font encoding

\usepackage[ruled]{algorithm2e}
\usepackage{amssymb,mathtools}
\usepackage{todonotes}
\usepackage{graphics}
\usepackage{graphicx}
\usepackage{subfig}
\usepackage{epstopdf}
\usepackage{colortbl}
\usepackage{multirow}
\usepackage{rotating}
\usepackage{pdflscape}
\usepackage{tabulary}
\usepackage{tabularx}
\usepackage{newtxtext}
\SetAlFnt{\algofont}
\SetAlCapFnt{\algofont}
\SetAlCapNameFnt{\algofont}
\SetAlCapHSkip{0pt}
\IncMargin{-\parindent}

\usepackage[framemethod=TikZ]{mdframed}
\mdfdefinestyle{SummaryFrame}{%
    linecolor=black,
    outerlinewidth=1pt,
    roundcorner=5pt,
    innertopmargin=\baselineskip,
    innerbottommargin=\baselineskip,
    innerrightmargin=10pt,
    innerleftmargin=10pt,
    backgroundcolor=gray!20!white
    }
\colorlet{tabgrey}{black!30}
\newcommand{\greyhline}{\arrayrulecolor{tabgrey}\hline\arrayrulecolor{black}}
\IEEEpubid{10.1109/COMST.2017.2782182~\copyright~2017 IEEE}
\begin{document}
%%%%%%%%%% Copyright page begin
\thispagestyle{empty}
\setcounter{page}{0}
\onecolumn
\textcopyright 2017 IEEE. Personal use of this material is permitted. Permission from IEEE must be obtained for all other uses, in any current or future media, including reprinting/republishing this material for advertising or promotional purposes, creating new collective works, for resale or redistribution to servers or lists, or reuse of any copyrighted component of this work in other works.
\nocite{8183438}
\twocolumn
\newpage
%%%%%%%%%% Copyright page end
% Title portion
\title{Simulating Opportunistic Networks:\\Survey and Future Directions}
\author{Jens Dede$^1$, Anna F\"orster$^1$, Enrique Hern\'{a}ndez-Orallo$^3$, Jorge Herrera-Tapia$^{3,4}$, Koojana Kuladinithi$^2$, Vishnupriya Kuppusamy$^1$, Pietro Manzoni$^3$, Anas bin Muslim$^1$, Asanga Udugama$^1$,   Zeynep Vatandas$^2$   \\
$^1$\textit{University of Bremen, Germany}\\
$^2$\textit{Hamburg University of Technology, Germany}\\
 $^3$\textit{Universitat Polit\`{e}cnica de Val\`{e}ncia, Spain}\\
 $^4$\textit{Universidad Laica Eloy Alfaro de Manab\'{i}, Ecuador}
}

\maketitle

\begin{abstract}
Simulation is one of the most powerful tools we have for evaluating the performance of Opportunistic Networks. In this survey, we focus on available tools and models, compare their performance and precision and experimentally show the scalability of different simulators. We also perform a gap analysis of state-of-the-art Opportunistic Network simulations and sketch out possible further development and lines of research.

This survey is targeted at students starting work and research in this area while also serving as a valuable source of information for experienced researchers.
\end{abstract}


\newcounter{TakeAwayMessageCounter}
\newcommand{\takeaway}[1] {
    \stepcounter{TakeAwayMessageCounter}
    \noindent
    \begin{minipage}{\columnwidth}%Prevent column break
       \vspace{2mm}
        \hspace{2ex}\textbf{Take-Away Message \arabic{TakeAwayMessageCounter}:}\newline
       \textit{#1}
       \vspace{2mm}
    \end{minipage}
  }

\IEEEpeerreviewmaketitle

% Head 1

% !TEX root = report-ieee.tex

\section{Introduction}

Interest in opportunistic networks (OppNets for short) has been growing in the research community ever since it first appeared. Their main function is to exploit direct, localized communications instead of infrastructure-based ones. This is also their main advantage: they can operate anywhere end-user devices are present. For this reason they are considered extremely useful for emergency scenarios, in rural or remote areas and for overloaded or sabotaged networks. They do not compete with mainstream 4G~/~5G networks: instead, they complement them.

The motivation for this survey stems mainly from the observation that opportunistic services and applications are still being deployed too slowly \cite{7823357}. In our opinion, one of the main reasons for this is the low level of readiness of the technologies being considered. However, in this case we think it is a problem of the 'chicken or the egg':
\begin{itemize}
    \item Researchers need large-scale deployments with real users, real devices and real scenarios to increase the robustness and reliability of the services being designed.
    \item Users do not want to use unreliable, research-level services.
\end{itemize}

Researchers must therefore depend on small-scale deployments and simulated environments to conduct their investigation. Consequently, we believe the best way to break out of this cycle is to have well-performing, realistic, reliable, and highly-scalable simulation tools.

This survey focuses on this very objective. It provides a thorough review of existing simulation models and tools available for opportunistic networks. We compare them in terms of their precision, scalability and performance. We focus on four large simulation tools that are free to use and widely accepted by the OppNets community, namely (in alphabetical order): Adyton, OMNeT++, ONE and ns-3.

There exist various relevant and thorough surveys on OppNets-related issues in the literature, one of the most widely used being the one by Mota et al.\ \cite{Mota:2014} which describes various protocols, mobility models and tools. In \cite{7080887}, the authors provide a detailed survey of available mobility models and study how mobility parameters affect performance.  Finally, in \cite{7056450,6619433,6231292,relpap3}, the authors focus on routing and data dissemination techniques, while in \cite{6977878} researchers concentrate more on the vehicular context.

\IEEEpubidadjcol %% Required to prevent an overrun in the copyright notice. Needs to be somewhere in the second column of the first page
However, when starting off research on this topic, the existing surveys and tutorials were of little help in answering some questions that arose. For example, \textit{"Should we use Poisson or constant traffic?", "Is it OK to use a trivial radio transmission model?", "Which tool should I use and when?", "Which models are available and where?", "What is the impact of individual models on the performance of the simulation tool?" or "Which tool scales better?"}.
In addition, we go beyond a traditional survey which simply lists and compares available options. We identify the gaps in the current state-of-the-art and propose possible future directions of the research in this area necessary to eventually reach our desired goal.

For this reason, we strongly believe that this survey provides a number of important guidelines for both students and experienced researchers wishing to explore OppNets. In the process of presenting these explanations, we have attempted to provide a simple description of each aspect (e.g., concepts, models, simulators, etc.) of OppNets so that a newcomer can easily follow and understand the content of this survey.

This survey continues as follows: the next Section~\ref{sec:def} defines what an opportunistic network (OppNet) is and narrows down the scope of the survey. Section~\ref{sec:perf} explores in detail available tools and methods for evaluating the performance of OppNets in terms of simulation, theoretical analysis, testbeds and real deployments. Section~\ref{sec:tools} presents the available tools (not models) capable of simulating OppNets behavior.
Section~\ref{sec:models} explores the relevant models for simulating OppNets, such as mobility, traffic, radio propagation, and many more.  Section~\ref{sec:data} briefly addresses the question of data propagation algorithms, which is typically the main research area in OppNets. However, a full overview of these algorithms and protocols is beyond the scope of this survey. Then, Section~\ref{sec:metrics} presents the evaluation metrics used for OppNets, before Section~\ref{sec:results} presents a study of the performance of the four simulation tools when compared to one another. Section~\ref{sec:best} summarizes the findings of this survey and offers a list of best practices. Finally, Section~\ref{sec:future} identifies gaps in current best practices and proposes some further concrete steps towards more scalable and realistic OppNets simulations and Section~\ref{sec:conc} concludes the paper.

% !TEX root = report-ieee.tex

\section{Opportunistic Networks: Definition and Concepts}\label{sec:def}

The term opportunistic networks has at times been used for different technologies. Here, 
we define Opportunistic Networks as the set of \textbf{applications and services running on end-user devices (e.g., smartphones, tablets and similar digital devices) that use \textit{direct communication opportunities} to exchange information with each other.} 

A typical sample application is a disaster alert, where an alert message (e.g. of a tsunami or earthquake) is sent immediately to all neighboring nodes. This way, all devices in the area can be easily reached, without need for a special application or paid mobile data subscriptions. Furthermore, if the communication infrastructure, such as 3G or 4G, is defective or non-existent, the messages still get exchanged. The term "opportunistic" refers to the fact that communication opportunities with other devices are used as they come.

There are many similar use-cases which exploit the OppNets concept, such as delay-tolerant networks (DTN), space communications, vehicle-to-vehicle communications and under-water communications. A good overview of such applications is provided by Mota et al.\ in \cite{Mota:2014}. Such networks and their relevant applications are closely related to OppNets, but their properties in terms of mobility and application requirements are quite different, which is why they will not be considered further here.

% !TEX root = ../report-ieee.tex

\begin{table}[htbp]
    \begin{center}
\caption{Various application examples used for motivating and evaluating OppNets.}
\label{tab:appl}
        \begin{tabular}{|p{1.5cm}|p{3cm}|p{3cm}|}
	\hline

        &\textbf{Destination-oriented (unicast or multicast)}
        &\textbf{Destination-free}\\
\hline
        \textbf{User-driven}&
        Announcements, messaging, chatting&
        Crowd sourcing, people as sensors, recommendations, public announcements\\
\hline
        \textbf{Automatic / event-based}&
        Security and safety, newsletters, health alerts (allergies, smoke, etc.)&
        Public safety alert (floods, strong sun radiation etc.)\\
\hline
        \textbf{Automatic / periodic}&
        Patient or baby monitoring, critical device status&
        Public reports (traffic, weather, health)\\
\hline
\end{tabular}
\end{center}
\end{table}

One additional aspect related to the use-cases under consideration is the classification of the applications used by users. Table~\ref{tab:appl} gives the main properties and examples of the various application types. We basically want to  differentiate between: (1) destination-oriented vs.\ destination-free models, and (2) user-driven vs.\ automatic applications. Destination-oriented applications assume that each message has a dedicated destination, whether that be a single user or group of users. Destination-less applications do not know this from the beginning - they assume all users might or might not be interested in their data. It is thus left to the propagation protocols to decide on where to deliver them.

This classification, though simple, is very important because it dictates the characterization of the data propagation to be used and the metrics to consider depending on the class of applications. Some protocols are designed to serve destination-less applications and will not perform well in destination-oriented applications, and vice versa.

% !TEX root = report-ieee.tex

\section{Performance Evaluation of Opportunistic Networks}\label{sec:perf}

Although this survey focuses on simulating methods, it is important to put it into context and briefly describe other methods for evaluating OppNets. Performance evaluation can be defined as quantifying the service delivered by a computer or a communication system \cite{Leboudec10}. In order to perform an evaluation, we must define first our evaluation goals, and then the system, load and metrics.

\begin{figure}[b]
  \centering
    \includegraphics[width=3.5 in]{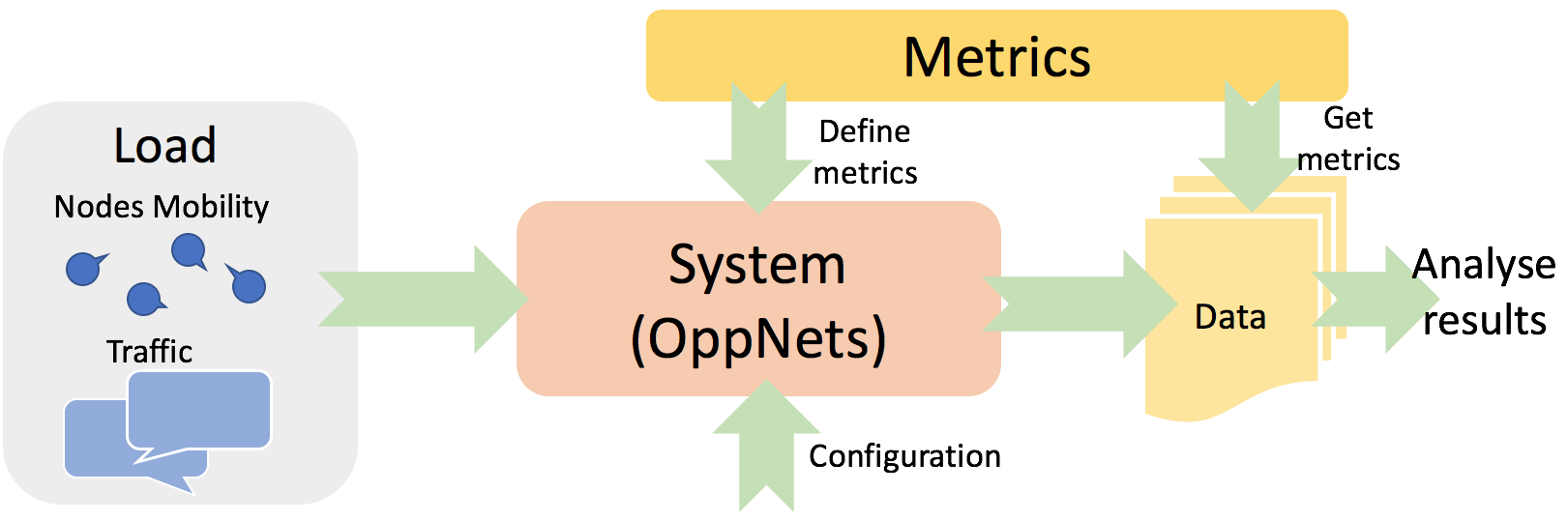}
  \caption{Flow diagram of the process for the performance evaluation of OppNets}
  \label{fig:performance_evaluation}
\end{figure}

First, we must define our goals. It may be obvious, but setting clear evaluation goals is critical to this evaluation process and will help in establishing the type of evaluation, defining the system and the load, and selecting the metrics to analyze. For example, the goal might be to compare different data propagation protocols, which means we must evaluate them under different scenarios and different loads in order to determine which protocol performs better in what context. On the other hand, if our goal is to design a new real-world application, then the evaluation must be performed using a load similar to the expected one in the real deployment scenario.

\begin{table*}[btp]
\begin{center}
\caption{Summary of the available performance techniques for OppNets.}
\label{tab:performance_methods}
\begin{tabulary}{1\linewidth\tymin=30pt\tymax=250pt}{|L|LLL|}
\hline
    & \textbf{Implementation (Real System)} & \textbf{Simulation} & \textbf{Analytical Model} \\

\hline
\textbf{System}  & Real & Depends on the simulator, from simple to complex & Simple. Strong assumptions and simplifications  \\
\hline
\textbf{Load} & Real or synthetic (system benchmarking) & Configurable: From simple loads (synthetic mobility models and simple scenarios) to realistic loads (trace-based node mobility and traffic) & Very simple (contact rate, simple area, no spatial consideration, single message)  \\

\hline
\textbf{Metrics}  & Experiment-bound & Custom metrics. Depends on the simulator, can be similar to real systems. & Deterministic values for population processes, and distributions for Markov Chain models \\
\hline
\textbf{Pros} & The most realistic method. & Very flexible: Full control of workload, scenario model, metrics, etc.\ with different resolution levels. Cost and time efficient & Fastest\\
\hline
\textbf{Cons}  & Very limited scenarios (it is not easy to evaluate the effect of varying parameters or real users). Very time and cost consuming  & Can deviate significantly from real systems. Can be very time-consuming for very large scenarios  &  Oversimplified results, considering ideal conditions only. Models very complex to obtain\\
\hline
\end{tabulary}
\end{center}

\end{table*}

The performance evaluation process consists of three main elements as shown in Figure~\ref{fig:performance_evaluation}. In detail, we have:
\begin{itemize}
  \item \textbf{System}. The system, in our case, is the opportunistic network to be evaluated and it comprises all elements from hardware to software that can affect the performance of these networks, such as communication range, data transmission speed, buffer capacity, overhead, user behaviour, etc.

    \item \textbf{Load}. The load (or workload) represents the type and quantity of requests in a system, which in OppNets are primarily mobile nodes with devices that transmit messages. More precisely, we can distinguish between:
    \begin{itemize}
    \item \textbf{Node mobility}: Defines the movement of the nodes, and their interactions are especially important for the evaluation of OppNets. Node movements are usually restricted to the \textbf{scenario} to be evaluated. The complexity of these scenarios can range from simple restricted square areas to realistic map-based scenarios. This topic will be further discussed in  Subsection~\ref{sec:mobility}.
    \item \textbf{Traffic load}: Refers to all the messages and network requests generated by the nodes. This traffic load can be as simple as the diffusion of one message to all the nodes in the network or to more complex messaging patterns that resemble the use of social or messaging mobile applications. Subsection~\ref{sec:traffic} further details traffic models.
    \end{itemize}
  \item \textbf{Metrics}: Finally, we also need to define the metrics for evaluating the performance of OppNets. These metrics focus mainly on evaluating information diffusion in terms of diffusion~/~dissemination time, delivery rate or number of hops. Regarding  nodes and network infrastructure utilization, we can also obtain metrics about buffer occupation, network overhead, energy consumption, etc. When the performance evaluation is completed, we usually have a large amount of data that we must process in order to get the desired metrics to proceed on to the analysis of the results. Regarding the statistical nature of the metrics, we can obtain deterministic values alone (for example, a mean), or we can determine its stochastic distribution. This will be discussed in more detail in Section~\ref{sec:metrics}.
\end{itemize}

In terms of how to perform such evaluations, we have the following options:
\begin{itemize}
    \item \textbf{Implementation} (real or testbed systems): Experiments performed using real scenarios and equipment can be very expensive and sometimes impossible to perform. Nevertheless, some complete evaluations in controlled places have been performed in \cite{Banerjee08,Vukadinovic11,Ristanovic12}. Other experiments are focused on obtaining traces about node mobility that can be used to simulate these scenarios, as shown in Subsection~\ref{sec:mobility}.
  \item \textbf{Simulation}: This is usually a simplified model of the system and the load implemented by software. A common approach is to combine a network simulation tool with realistic mobility traces in order to reproduce the real dynamics and interactions of mobile nodes. Nevertheless, as we will see in this survey, simulations can be computationally intensive (especially for large or complex loads) and its parametrization is not trivial.
  \item \textbf{Analytical Model}: This is a mathematical model of the system and the load. Analytical models can avoid some of the drawbacks of simulations and real testbeds by providing a faster and broader performance evaluation, one in which the key mechanisms underlying the information diffusion can be identified. Analytical models usually require strong assumptions or simplifications about the system to be evaluated, and the load model considered is very simple. Usually, the diffusion of a single message in a network of nodes with a given contact pattern is Poisson-distributed (considering that the inter-contact times distribution between pairs of nodes is exponentially distributed with a given contact rate) \cite{Groenevelt05}. Despite this, it is not trivial to obtain such models. Two main classes of analytical models have been proposed for modeling such network dynamics: deterministic models based on population or epidemic processes \cite{Haas06,Zhang07,DeAbreu14,Xu15,Hernandez16} and Markovian models \cite{Groenevelt05,Haas06,Spyropoulos08,Hernandez15,Karaliopoulos09,Whitbeck11}. Markovian models, being stochastic, have the benefit of obtaining the probability distribution of the metrics at the cost of an increased computational cost when no closed-form expressions can be obtained.

\end{itemize}

\begin{figure}[tb]
  \centering
    \includegraphics[width=\columnwidth]{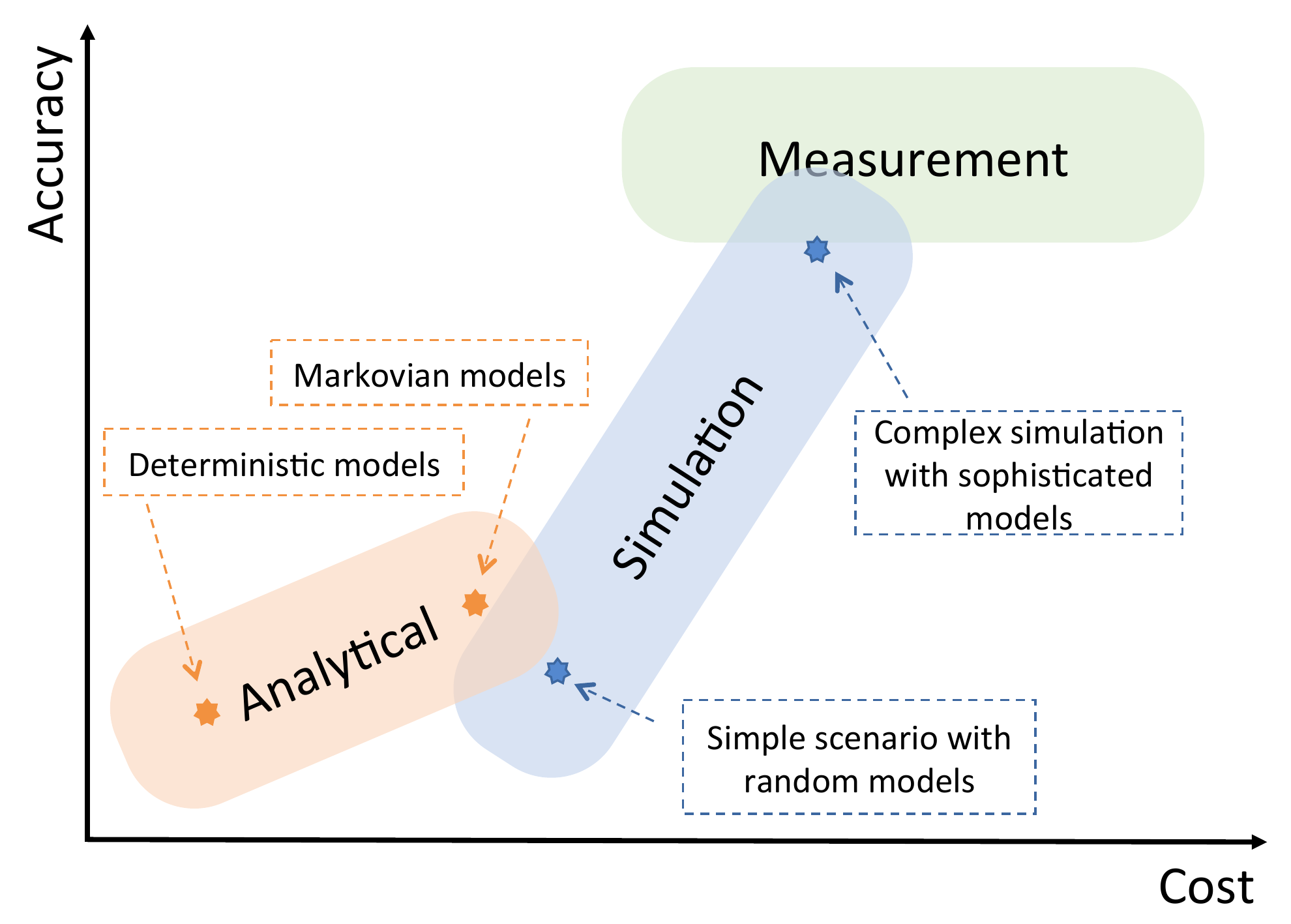}
  \caption{Accuracy versus cost of the different methods of performance evaluation.}
  \label{fig:precision_of_methods}
\end{figure}

The main performance evaluation methods are summarized in Table~\ref{tab:performance_methods}. One of the most significant aspects of selecting the type of evaluating method is its precision to cost ratio. In Figure~\ref{fig:precision_of_methods} we can see the relationship between these two factors. On one side, we have the cost, used as a broader term, comprising both the time cost and the monetary cost (which can be very high in real experiments). On the other side, we have the accuracy of the obtained results compared to the real-world deployments. Analytical methods are very fast, but the results can be unrealistic. Simulation can provide precision very close to real testbeds using sophisticated simulation models, despite its computational cost. Thus, in this survey we focus on simulation models and how to obtain the most realistic and best performing scenarios. 

% !TEX root = report-ieee.tex

\section{Simulation Tools}\label{sec:tools}

In this section we first give an overview of existing simulation tools and platforms suitable for evaluating OppNets. While there exist many other simulation tools, we focus on those which are free to use and widely accepted. 
We explore them in terms of platforms supported, available OppNets models and general advantages~/~disadvantages. We further explore them experimentally in Section~\ref{sec:results}, where we provide a direct comparison.

\subsection{OMNeT++}\label{sec:tools-omnet}

OMNeT++\footnote{https://omnetpp.org} is an extensible, object-oriented, general purpose discrete event simulator (DES) written in C++. It was first published in late 1990 and its current version is 5.0. OMNeT++ has a free academic license and provides the mechanisms for building and simulating any type of network, from sickness dissemination to wireless networks. The OMNeT++ simulator itself only provides the blocks for building nodes in networks and modeling their behavior and interactions. So called frameworks have emerged over the years, that make available pre-defined nodes with some special behavior: sensor nodes, IP-based nodes, satellites, cars, etc. INET\footnote{https://inet.omnetpp.org} is one such framework that implements the protocols of the Internet Protocol (IP) suite. Using INET, users can easily construct a network of IP-based nodes and focus on parameters, applications, scenarios, etc. OMNeT++ is especially well suited for such broad studies as it offers a user-friendly environment for automatic parameter studies.

\subsubsection{Available Models for OppNets}

INET itself is not very well suited directly for OppNets simulations as it focuses on IP-based networks and their protocol stack. This stack is overly-complex for efficient OppNets simulations, as we will also identify later throughout this survey. In addition, INET does not readily provide OppNets data dissemination protocols. Some of the authors of this survey have used INET as a basis for our OPS framework\footnote{https://github.com/ComNets-Bremen/OPS}. It offers the possibility of switching the IP stack of INET on and off and currently offers some simple data dissemination protocols. This is also the framework that we use in this survey with the latest OMNeT++ 5.0 version (2016) and INET Framework version 3.4.0 (2016).

Additionally, other frameworks and models have also been developed for OppNets in OMNeT++. However, all these models have been developed for earlier versions of OMNeT++ (mainly 4.1-4.2, 2010-2013) and need large-scale refactoring. Such frameworks include OPPONET~\cite{OPPONET,OPPONETMIXIMEXT,OPPONETCACHE,OPPONETFRAMEWORK} and OppSim~\cite{OPPSIM}.

\subsubsection{Advantages}

OMNeT++ has a very sophisticated user interface, with many possibilities for visualizing and inspecting various scenarios. The latest version also includes a sophisticated 2D and 3D graphics support for visualizing scenarios. It is also highly modular while also clearly separating node behaviour from node parameters, making it very easy to run large parameter studies. Its performance is also very good; it runs on all major operating systems and has very well maintained documentation, including user guides, tutorials, wiki page, etc. It can also be parallelized.

\subsubsection{Limitations}

OMNeT++ does not offer the possibility of transferring simulation models to real implementations, e.g.\ directly to Linux distributions.

\subsection{ns-3}
Network Simulator 3 (ns-3)\footnote{https://www.nsnam.org} is a discrete event simulator primarily focused on simulating IP-based networks with an emphasis on the network layer (layer 3) and the above layers of the protocol stack. The simulation scenarios can be created using C++ or Python, and they are run as command line applications with no GUI. However, \textit{NetAnim} is a tool that ships with ns-3 for visualizing node mobility during or after a simulation. ns-3 is licensed under the GPLv2 open source license. ns-3 was completely rewritten, although it can be considered to be the successor to ns-2. For this reason, ns-3 is incompatible with ns-2 and the simulation models have to be adapted to be used in ns-3. At present, ns-2 is only lightly maintained as the focus of the developers is on ns-3, which should therefore be used for new projects. The version used in the experiments is ns-3.27.

\subsubsection{Available Models for OppNets}
ns-3, like OMNeT++, is a general purpose network simulator. In order to be used for OppNets simulations, it needs to be configured with the proper protocols at all levels---e.g.\ a OppNets data propagation protocol, mobility, traffic, link technology, etc. It is currently not possible to switch off the link technology model.

\subsubsection{Advantages}
All communication modules have been designed in such a way as to easily match the interfaces to the standard Linux APIs. For this reason, it is possible to easily move a simulation setup to the real world and vice versa. It is also possible to interact with existing real-world networks using virtual TAP (layer 2) devices, connecting simulated and real nodes.

\subsubsection{Limitations}
A good user interface with debugging and visualization options is clearly missing. Furthermore, the simulator structure is rather complex and not easy to parametrize or change, even for experienced researchers.

\subsection{The One}

The ONE (Opportunistic Network Environment) \cite{ONE:2009}, is a simulation tool designed specifically for OppNets. It was first developed in 2009 at Aalto University, and it is now maintained cooperatively by Aalto University and the Technische Universit\"at M\"unchen. The version used in this survey is 1.6.0, released in July 2016, and is available on GitHub\footnote{https://akeranen.github.io/the-one}. It is written in Java.

\subsubsection{Available Models for OppNets}

The ONE is designed specifically for OppNets and therefore offers a wide variety of models that continues to grow.
The ONE allows for the generation of node movements using different models, the reproduction of message traffic and routing, cache handling and the visualization of both mobility and message passing through its graphical user interface.
It can also produce a variety of reports, such as node movements to message passing and general statistics.

\subsubsection{Advantages}
In terms of available models for OppNets, the ONE is very sophisticated. It is easy to use and has a solid user community.

\subsubsection{Limitations}
The ONE does not perform well for large simulations, mainly because of the programming language. In terms of models, it does not offer any link technology models nor radio propagation models.

\subsection{Adyton}
\label{sec:tools:adyton}

Adyton~\cite{Adyton} is the newest tool in our survey, released in 2015. It is written in C++ and is also dedicated to simulating OppNets scenarios. It is only available on Linux and does not have a graphical user interface. It is freely available on Github\footnote{https://npapanik.github.io/Adyton}.

\subsubsection{Available Models for OppNets}
As is the ONE, Adyton was designed specifically for OppNets. It has a wide variety of data propagation models, while also offering buffer and cache size management. In terms of mobility, Adyton took a different path from the other simulators described here. Instead of actually moving nodes around, it pre-calculates contacts between them.
This is a very processing-efficient way, but only a few mobility traces are made available by the developers. To use other traces, they require pre-processing by an external tool.

\subsubsection{Advantages}
The performance of Adyton is very promising and it seems to solve the main problem of the ONE. In terms of available models, it also offers an ever-growing variety of data propagation models.

\subsubsection{Limitations}
A good graphical user interface is missing for debugging and validation. In addition, the availability of mobility models needs to be improved to allow for more independent studies. It does not support link technologies, radio propagation models nor energy consumption models.

\subsection{Further tools}
\label{sec:further_tools}

In addition to the simulation tools described above, there are some other mobility-oriented tools that are also very useful for simulating OppNets.

\begin{itemize}
  \item \textbf{BonnMotion} is a tool than can create and analyze mobility scenarios and is most commonly used for researching mobile ad hoc network characteristics \cite{bonnmotion}. BonnMotion's main objective is to create mobility traces using mobility models, as well as trace-based mobility scenarios (both described in Subsection~\ref{sec:mobility}). Furthermore, the generated traces can be exported to a notable number of formats used by network simulators, such as \textit{ns-3}, \textit{Cooja}, \textit{The ONE}, \textit{OMNeT++}, \textit{GloMoSim~/~QualNet} etc.

  \item \textbf{Legion} is a commercial tool for generating mobility scenarios used by architects and civil engineers for urban and traffic planning. Using this mobility simulator, we can create our own scenarios (mainly buildings or city areas) and define the number of pedestrians~/~vehicles, their type of movement and destination. The underlying model is patented and validated with large traces of real pedestrian movement. The output generated by these tools can be used as input to an OppNets simulator after re-formatting. A good example is described in \cite{Helgason14}.
   \item   \textbf{PedSim} (Pedestrian Simulator) \cite{PedSim} is similar to Legion, but free to use.
  \item \textbf{SUMO} (Simulation of Urban MObility) is also a mobility traffic generator written by the German Aerospace Center (DLR) \cite{sumo}. The main focus is on the simulation of public transport, pedestrians and vehicles, including speed limits, traffic lights etc. From these simulations, mobility traces can be exported for trace-driven mobility in OppNets simulations.
\end{itemize}

Summing up, the mobility-generating tools presented here can be very helpful in relieving the OppNets simulator from some of its tasks and in producing very sophisticated large-scale mobility traces.

 %!TEX root = report-ieee.tex

\section{Simulation Models}\label{sec:models}

The evaluation of opportunistic networks involves various different elements that together make simulation results more realistic. In this section, we focus on the most relevant of these elements by presenting the available models and their properties. Though this may not be an exhaustive survey, the next subsections do provide an executive summary of the state-of-the-art models for mobility, user behaviour, radio transmission and interference, battery and energy consumption, traffic generation and link technologies.

\begin{figure*}[!t]
    \centering
    \includegraphics[width = \textwidth]{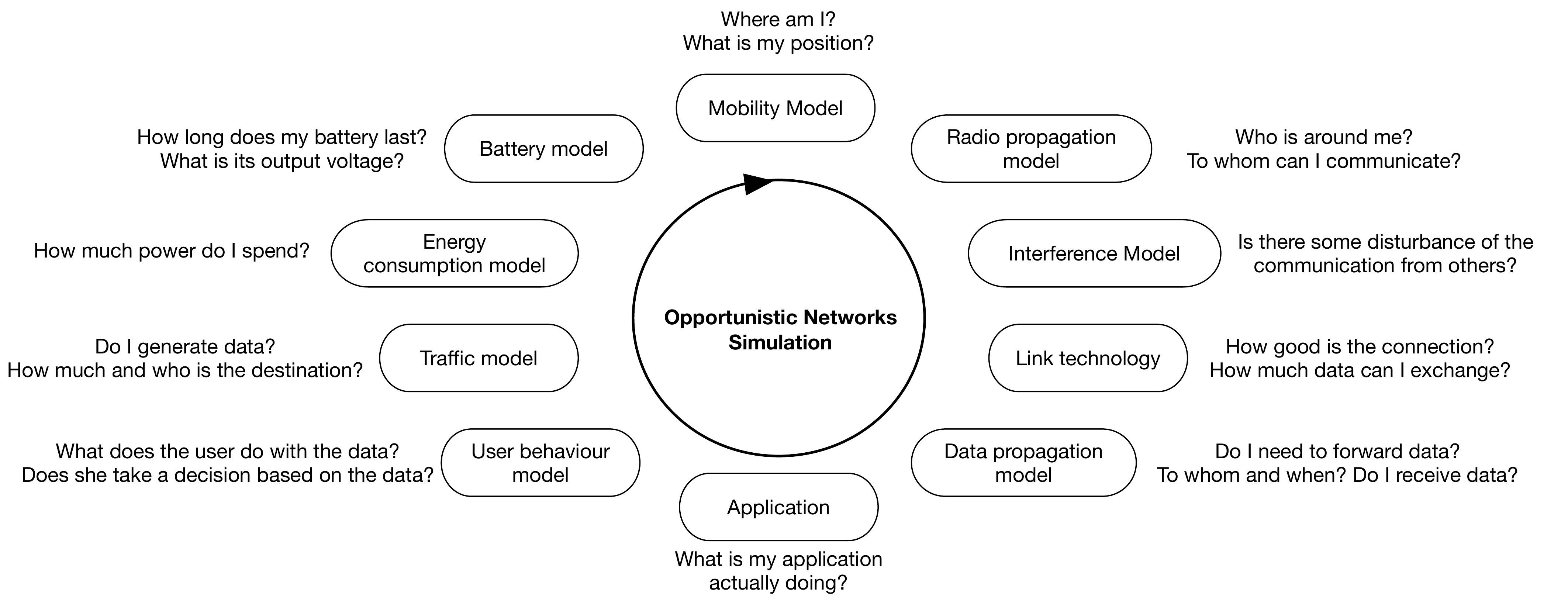}
    \caption{The OppNets simulation models explored in this survey from the perspective of a single simulated node. }
    \label{fig:models}
\end{figure*}

Figure~\ref{fig:models} illustrates the main purpose of the individual models explored here. The models are presented from the perspective of a single simulated node.

% !TEX root = report-ieee.tex

\subsection{Mobility Models}\label{sec:mobility}

\begin{figure}[htb]
\centering
\includegraphics[width = 0.9\columnwidth]{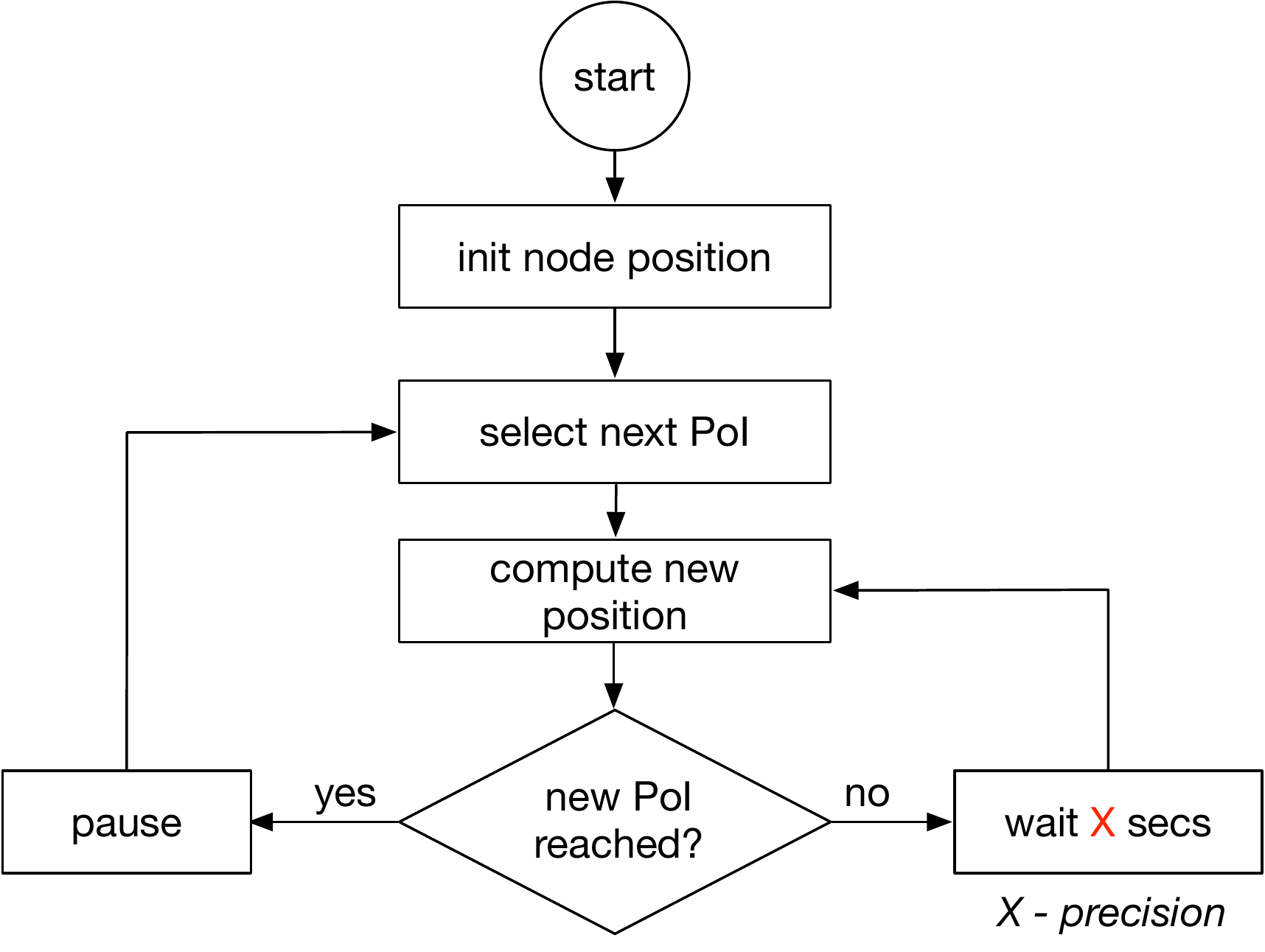}
\caption{The general structure of a  mobility algorithm. }
\label{fig:str-mob-models}
\end{figure}

This subsection focuses on how the movement of the users is modelled in simulators. Mobility models are designed to describe the movement patterns by using the user's location and velocity change over time.
Since mobility provides opportunities for contacts and therefore communication, understanding human movement patterns is an essential element for evaluating a protocol performance. It is desirable for mobility models to emulate the
movement pattern of targeted real life applications in the most realistic way possible.
The general structure of a  mobility algorithm  typically follows the steps indicated in Figure~\ref{fig:str-mob-models}.

Mobility models can be categorized into three main types based on how the next \textit{Point of Interest (PoI)} is selected.

\paragraph {\textbf{Random Mobility Models}} Models with random movements employ stochastic movement patterns to move a node within a given area. In many simulations, the Random Way Point (RWP) model is the most frequently used mobility model due to its simplicity of implementation. The mobile node selects a destination randomly and moves towards it with a randomly selected speed. The other extended mobility models based on RWP can be further characterized as mobility models with  \textit{temporal dependency} (i.e., the next movement will depend on the previous history), with  \textit{spatial dependency} (i.e., tend to change the movement in a correlated manner) and with  \textit{geographic restriction} (i.e., movement is restricted by streets or obstacles). Almost all the available random mobility models are discussed in detail in \cite{baisurvey} and \cite{camp2002survey}.

These models are not very useful for evaluating OppNets as the mobility patterns used in OppNets should mimic human behavior, in which a user's movement is also influenced by their daily activities (at home, school or work), by their means of transportation (walking, bike, bus, etc.) or by the behavior of other users in their neighbourhood~/~environment. Many of these activities are related to user behaviour and social relationships, in addition to the three characteristics mentioned above. We will show these properties and their impact on OppNets performance experimentally at the end of this subsection as well.

\paragraph {\textbf{Real Mobility traces}} There is a large collection of datasets obtained from the observation of nodes mobility in real scenarios. These traces provide accurate information, especially when the trace is being collected among a large number of participants and for a longer period of time. We can find two classes of traces: contact-based and location-based. A contact-based trace is obtained by measuring the times contacts between pairs of nodes occur for a given time interval. Well-known datasets such as \textit{Infocom} \cite{Hui05}, \textit{Cambridge} \cite{Leguay06}, \textit{Milano} \cite{Gaito09}, \textit{MIT (or Reality)}  \cite{Eagle05} among others, have been extensively used, and their statistical properties have been studied in depth \cite{Passarella11,Hernandez16b}. The simplicity of these traces allows for the analytical evaluation and simple simulation of OppNets.

 Their main drawback is that they do not allow the impact of communication protocols to be simulated. Therefore, to evaluate these aspects, we must use location-based traces. These traces are the result of obtaining the location of the nodes (mainly GPS coordinates) periodically (or when they move). There are several types of traces, such as taxi mobility in Shanghai \cite{Zhu10}, or student mobility on the National Chengchi University campus \cite{nccu_traces}, among others. The Crawdad repository \cite{Crawdad} contains most of the publicly available traces, and it might be considered the first place to look for required  traces for simulations.

 Though trace-based mobility models represent very realistic movement patterns, they may not be very useful when it comes to new scenarios in which the collection of traces has not yet been carried out. Furthermore, they are fixed, i.e., they cannot be scaled up or extended in time. Also, they cannot react to any changes from the user's perspective, e.g.\ modelling the user running away from an unexpected, dangerous situation. They are also expensive to use in simulations because the information about the next points has to be read from external files. We will demonstrate this at the end of this subsection.

\bgroup
\def\arraystretch{1.5}
\begin{table*}[bhtp]
\caption{A comparison between the different hybrid mobility models for OppNets (RWP is used here only as a reference).}

    \setlength{\tymin}{8em} % Set minimum table width to overcome issue with ugly line breaks
\begin{tabularx}{\textwidth}{|X|c|ccccccccc|}
\hline
    \textbf{Property / Model}&
    \rotatebox[origin=c]{90}{\textbf{RWP}}&
    \rotatebox[origin=c]{90}{\textbf{SLAW}}&
    \rotatebox[origin=c]{90}{\textbf{SMOOTH}}&
    \rotatebox[origin=c]{90}{\textbf{TLW}}&
    \rotatebox[origin=c]{90}{\textbf{HCMM}}&
    \rotatebox[origin=c]{90}{\textbf{TVC}}&
    \rotatebox[origin=c]{90}{\textbf{SOLAR}}&
    \rotatebox[origin=c]{90}{\textbf{SWIM}}&
    \rotatebox[origin=c]{90}{\textbf{HHW}}&
    \rotatebox[origin=c]{90}{\textbf{HWDM}}\\
\hline
    \textbf{Distributions of PoIs (Point of Interests) are}&&&&&&&&&&\\
\greyhline
    Random&\checkmark&-&-&\checkmark&\checkmark&-&\checkmark&-&\checkmark&\checkmark\\
\greyhline
    Based on Truncated Power Law (TPL) distribution&-&-&\checkmark&-&-&-&-&-&-&-\\
\greyhline
    Based on traces / maps -- without context&-&\checkmark&-&-&-&\checkmark&-&\checkmark&-&\checkmark\\
\greyhline
    Based on traces / maps -- with context (e.g.\ home, work, supermarket)&-&\checkmark&\checkmark&-&-&\checkmark&\checkmark&\checkmark&\checkmark&-\\
\hline
    \textbf{Next PoI selection based on\ldots}&&&&&&&&&&\\
\greyhline
    Random&\checkmark&-&-&\checkmark&-&-&\checkmark&-&\checkmark&\checkmark\\
\greyhline
    Based on previously visited PoI&-&-&\checkmark&-&-&-&-&-&-&-\\
\greyhline
    Distance from last location&-&\checkmark&-&-&-&-&-&\checkmark&-&-\\
\greyhline
    Time of day / week&-&-&-&-&-&-&-&-&\checkmark&\checkmark\\
\greyhline
    Based on social relationship&-&-&-&-&\checkmark&-&-&-&-&-\\
\greyhline
    Traces&-&-&\checkmark&-&-&\checkmark&-&\checkmark&-&-\\
\greyhline
    Vicinity of home location&-&-&-&-&-&-&-&\checkmark&-&-\\
\greyhline
    Popularity of places in general&-&-&-&-&-&-&\checkmark&\checkmark&\checkmark&-\\
\greyhline
    Popularity among friends&-&-&-&-&-&\checkmark&-&-&\checkmark&-\\
\greyhline
    Move with a community (e.g.\ friends)&-&-&-&-&\checkmark&-&-&-&-&\checkmark\\
\greyhline
    Differentiation between communities (e.g.\ friends, family, collogues)&-&-&-&-&-&-&-&-&\checkmark&-\\
\greyhline
    Personal preferences / contextual information (visit often cinemas but no restaurants)&-&-&-&-&-&\checkmark&\checkmark&-&-&-\\
\greyhline
    Level of personal mobility (some people move a lot, some doesn't)&-&-&-&-&-&-&\checkmark &-&-&-\\
\hline
    \textbf{Stay in a PoI for\ldots}&&&&&&&&&&\\
\greyhline
    Random time&\checkmark&-&-&\checkmark&\checkmark&\checkmark&\checkmark&-&-&\checkmark \textsuperscript{a}\\
\greyhline
    Based on TPL&-&-&\checkmark&-&-&-&-&-&-&-\\
\greyhline
    Trace-driven&-&\checkmark&\checkmark&-&-&-&-&\checkmark&-&-\\
\greyhline
    Course grained context (e.g.\ home: 16 hrs, work: 8 hrs)&-&-&-&-&-&-&-&-&\checkmark&\checkmark\\
\greyhline
    Fine grained context (e.g.\ supermarket: 30 min, bar: 1 hr, concert: 3 hrs)&-&-&-&-&-&-&-&-&-&-\\
\hline
    \textbf{Type of movements between PoIs\ldots}&&&&&&&&&&\\
\greyhline
    In a straight line with the same speed&\checkmark&\checkmark&\checkmark&\checkmark&\checkmark&\checkmark&\checkmark&\checkmark&\checkmark&\checkmark\\
\greyhline
    With different modes of transportation (different speeds)&-&-&-&-&-&-&-&-&-&\checkmark\\
\greyhline
    Map-driven (e.g.\ Navigator like)&-&-&-&-&-&-&-&-&-&\checkmark\\
\greyhline
    Trace-driven (e.g.\ GPS traces)&-&-&-&-&-&-&-&-&-&-\\
\greyhline
    Depends on real traffic -- day / night (Google Maps like)&-&-&-&-&-&-&-&-&-&-\\
\greyhline
    Group-based (e.g.\ move together with friends)&-&-&-&-&\checkmark&-&-&-&-&\checkmark\\
\hline
    \textbf{Allow scheduling / planning}&&&&&&&&&&\\
\greyhline
    Handle influences of user behaviour&-&-&-&-&-&-&-&-&-&-\\
\hline
    \multicolumn{11}{l}{\textsuperscript{a} defined in settings}\\
\end{tabularx}

\label{tab:syn-trace-models}
\end{table*}
\egroup

\paragraph{\textbf{Hybrid Mobility Models}}
This category shows a combination of the first two described above. In these models, some parameters (e.g., frequency of user movements w.r.t. locations) for a random model are derived based on a collection of traces or based on user experience. For example, the Small Worlds In Motion (SWIM) model \cite{SWIM} is based on the assumption that a user either selects a location close to her home or a very popular location (e.g.\ a popular restaurant in town) as a next PoI.
Thus, hybrid models attempt to model real properties of human movements by taking into account not only ``common sense'' assumptions, but also statistics from traces. Hybrid models achieve better performance and scalability compared to real mobility traces.

There are many studies that aim at providing realistic mobility models based on a social relationship (e.g., users periodic travels over short distances, movements coordinated by social relationships, etc.) between users and location preferences. Some of these models are: TLW (Truncated Levy Walk) \cite{TLW}, SLAW (Self-similar Least Action Walk) \cite{SLAW}, SMOOTH \cite{SMOOTH}, SWIM \cite{SWIM}, HCMM \cite{HCMM}, WDM (Working Day Model) \cite{WorkingDayModel}, TVC (Time Variant Community model) \cite{TVC},  HHW \cite{HHW}  and SOLAR (Sociological Orbit models) \cite{SOLAR}.

Unlike other surveys (such as \cite{karamshuk2011human,Pirozmand14}), in Table~\ref{tab:syn-trace-models} we offer a comparison of these models according to the human mobility properties they model, a more convenient approach when dealing with OppNets. We have organized the table according to the steps shown in Figure~\ref{fig:str-mob-models}.

We first explored how PoIs are identified by answering the following questions: Are those coordinates random? Are the coordinates taken from maps~/~traces, but are we missing any context information about them (e.g., a cafe or a concert hall)? Or, is all this information available?

Next, we explored how the next PoI is selected. Is this done purely randomly, or might we consider some human behaviour properties? For example, it could depend on the time-of-the-day, on previously visited locations (we tend to re-visit locations), but also on how much somebody likes going out.

Then we considered the pause time in a particular PoI. Again, random time can be assumed or the time spent in a location could depend on its specific type (e.g., only 10-30 minutes in a supermarket, but 2-3 hours in a cinema).

Finally, we considered how the user moves between PoIs. Is it on a straight line with constant speed or is it more realistic, e.g.\ sometimes taking the bus, sometimes walking and even a combination of those? It is interesting to note the difference between the ``Map-driven'' line with respect to the  ``Trace-driven'' and ``Depends on real traffic''. In defining the exact path, the last two are more precise considering the fact that daytime traffic conditions could be different (using information provided, for example, by Google Traffic) or, even more realistic, it could be taken from real traces with real people (who in general do not behave predictably).

\paragraph{\textbf{Performance comparison}}
To compare the performance of the different mobility models in terms of both the simulation time needed (i.e.\ the wall clock time) as well as in terms of other OppNets metrics, we designed a set of experiments whose results are presented in Table~\ref{tab:mobility-results}. 
% !TEX root = ../report-ieee.tex

\begin{table}[pt]
    \begin{center}
    \caption{A comparison between a simple random, real and hybrid mobility models in OMNeT++. The network consists of 50 nodes.}
     \label{tab:mobility-results}
        \begin{tabular}{|c||c|c|c|}

        \hline
        \textbf{Model} & \textbf{RWP} & \textbf{SWIM} & \textbf{SFO trace} \\
        \hline
        \hline
        \textbf{Simulation Time} & 4 min & 59 min & 109 min \\
        \hline
         \textbf{Memory used} & 74 MB & 86 MB & 127 MB\\
        \hline
        \hline
        \textbf{Average delivery rate} & 3 \%&  96\% & 92 \% \\
        \hline
        \textbf{Average delivery delay} & 20.6 h & 16.25 h & 13.16 h\\
        \hline
        \hline
       \textbf{Total number of contacts} & 190 & 46,752 & 155,757 \\
        \hline
        \textbf{Average contact duration} & 117.14 sec& 150.12 sec & 584.39 sec \\
        \hline
               \end{tabular}
    \end{center}
\end{table}%

For this, we took a set of real traces from San Francisco taxis~\cite{SFO_Cabs2009} (also referred to as "\textit{SFO trace}"; more details about these traces are given in Section~\ref{sec:results}) and evaluated them statistically in terms of covered area, number of nodes, duration of trace, mean speed of the nodes, etc. Then we parametrized a random model (RWP) and a hybrid model (SWIM) as closely as possible to the real trace. In order to keep the simulation duration viable, we used only 50 nodes.

From these results, we can conclude that real mobility traces are very time consuming to use, as the simulation took more than 20 times longer to complete compared to the corresponding simulation using a random mobility model. The hybrid model, in this case SWIM, positions itself between the two extremes. However, when comparing the use of other specific OppNets performance metrics, we see that RWP delivers completely different results from those of the real trace, which we assume to be the most correct one. Thus, its behaviour is obviously too far away from real movements and should not be used for OppNets simulations. SWIM delivers results similar to the real trace results in terms of delivery rate and delay, but looking into the contacts information, it becomes obvious that it is quite different in its behaviour. This, however, is due to the fact that SWIM is modelling walking users behaviour, while the real trace delivers information about users in vehicles (taxis).

\takeaway{Random mobility models are not well suited for simulating OppNets. One should preferably use sophisticated hybrid models, but if these models are inadequate for the scenarios being evaluated, consider using real traces. }

What we miss in all available models is the possibility of \textbf{reacting} to some message, such as running away from a fire, going to a party, etc. We will discuss this in Section~\ref{sec:future}, where we summarize all future ideas and lacking functionality.

% !TEX root = report-ieee.tex

\subsection{Radio Propagation Models}

Radio propagation models simulate how the wireless signal propagates through different environments, over different distances and obstacles. Since all opportunistic networks are typically wireless, these models are very relevant. At the same time, there is a trade-off between the precision of the radio propagation models against simulation time (i.e.\ wall clock time) and scalability of simulations. A more sophisticated radio propagation model will more correctly identify impossible or interrupted transmissions. However, when we are interested in the general behaviour of very large OppNets, a sophisticated radio propagation model would probably not significantly contribute to the understanding or evaluation of the system, while also using up too many resources.

Much like mobility models, radio propagation models (radio models for short) can be divided into trace-based and synthetic ones, plus some hybrid variants.

\subsubsection{Synthetic Radio Propagation Models}

Synthetic radio models attempt to mathematically describe the propagation of the radio signal through the environment and predict its strength at the receiver. In general, they calculate the so-called \textbf{Path Loss}, that is, the loss of strength between the sender and the receiver, as:

\begin{equation}\label{eq:pl}
P_R = P_T + G_T + G_R - PL - L_T - L_R
\end{equation}

\noindent
where $PL$ is the path loss, $P_T$ and $P_R$ are the transmitted and the received power, $G_T$ and $G_R$ are the gains of the transmitting and the receiving antennas, and $L_T$ and $L_R$ are the losses of the transmitting and receiving systems (e.g.\ coax, connectors, etc.). All parameters are in dB. This equation is often called also the \textbf{link budget} equation~\cite{rappa}. Very often, the system losses get neglected in simulation models and only the antenna gains are used.

The path loss is due to various factors, such as:

\begin{itemize}
\item Spreading: When the signal is emitted from the sender, it does not travel in a line, but rather spreads uniformly around it. This, of course, also happens in vacuum. The signal gets weaker with distance according to the inverse square law.
\item Attenuation: When the signal loses some of its power due to interaction with the medium. In a perfect vacuum, the attenuation should be zero, but a perfect vacuum does not exist, even in space.
\item Fading: The attenuation itself changes over time and~/~or distance.
\item Doppler effect: When the sender is mobile, the signal gets stretched or contracted. This impacts the frequency of the signal as it travels and is a specific kind of fading.
\item Shadowing~/~Superposition: When the signal interacts with other electromagnetic signals or reflections of itself. This is often also referred to as interference. For OppNets, relevant interference sources are all devices working in the same frequency bands as OppNets link technologies (mostly 2.4 GHz) and some other more unexpected devices, such as microwaves.
\end{itemize}

Most often, models use a more or less sophisticated version of the following equation:

\begin{equation}\label{eq:pl_det}
PL(d)[dB] = \overline{PL}(d)[dB] + X_\sigma [dB]
\end{equation}

\noindent
where $\overline{PL}(d)$ is the mean path loss for some distance $d$ from the transmitter and $X_\sigma$ is a probability distribution, typically Gaussian with zero mean and standard deviation $\sigma$. This distribution function models the above mentioned factors as \textit{noise} in the system.

\begin{figure}[tbhp]
\begin{center}
\includegraphics[width = 2in]{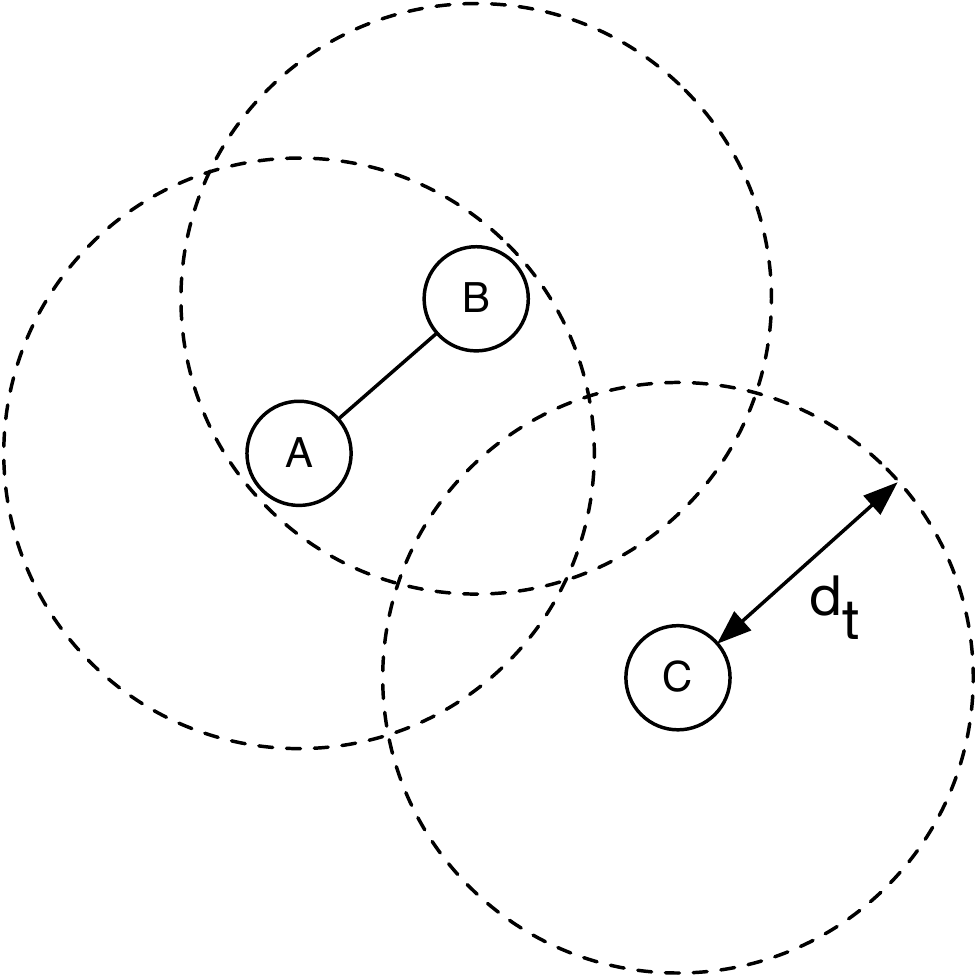}
\caption{The Unit Disk Graph model.}
\label{fig:udg}
\end{center}
\end{figure}

Models usually target either the calculation of the path loss (deterministic models) or the calculation of the noise (non-deterministic models). Reality is mirrored best with combined methods. However, most of the path loss models can be easily combined with the noise models, as it is a simple sum (see Equation~\eqref{eq:pl_det}).

The simplest model is the so-called \textbf{Unit Disk Graph (UDG)}. The idea is shown in Figure~\ref{fig:udg}: two nodes in a network are connected to each other if the distance between them is below some threshold $d_t$. In this case, the path loss is assumed to be zero. Otherwise, they cannot communicate to each other at all and their path loss is considered to be some maximum value, which makes the transmission impossible. Often this model is also called the \textbf{Binary model}, since it either allows for perfect communication or none at all. Obviously, this model is not very realistic, though it is very efficient.

Looking back at our Equation~\eqref{eq:pl_det}, UDG can be expressed as:

\begin{equation}\label{eq:udg}
PL(d)[dB] =
\begin{cases}
0, & \text{if } d \leq d_t \\
\infty, & \text{otherwise}
\end{cases}
\end{equation}

\noindent
where the noise is always 0. This can be easily extended to also include noise and to make the borders of the unit disk fuzzier.
The noise distribution can be Gaussian with zero mean and some standard deviation $\sigma$, but it can also be some other distribution.

Typically, the noise is considered to be independent spatially and timely. In other words, every time you need to calculate the path loss at some destination point, you do so independently of what you calculate at other positions (spatial independence) or for the same position at a different time (timely independence). However, many researchers have shown that this is not the behaviour of real world wireless links, so both dimensions (time and position) are highly relevant.

Some researchers have considered the spatial dependence in ~\cite{zhou:2006}. They have shown that the noise at a particular angle is dependent on the noise at neighbouring angles, which is defined by Equation~\eqref{eq:rim}. This model is referred to as the \textbf{Radio Irregularity Model (RIM)}:

\begin{equation}\label{eq:rim}
\begin{split}
PL^{DOI}_d = PL_d \cdot K_i, \\
\text{where }\\
K_i =
\begin{cases}
1, & \text{if } i = 0 \\
K_{i-1}\pm rand \cdot DOI, & \text{if } 0 < i < 360
\end{cases}
\end{split}
\end{equation}

In other words, instead of adding noise to the path loss independently of the position of the destination, it is calculated dependent on the noise for destinations close by. The DOI (degree of irregularity) is a parameter of how strong the noise is. The larger the DOI, the more irregular is the transmission area around the sender. A DOI of 0 results in a UDG model. The random distribution applied is the Weibull distribution, which is often used to model natural phenomena~\cite{zhou:2006}. Any PL model can be easily plugged into Equation~\ref{eq:rim}.

The second problem, the timely dependence, has been described in \cite{levis07beta}. There, the authors have explored the property of link burstiness. This means that if the communication between two nodes fails at some point, the probability that it will fail again grows, and the opposite. The authors have defined for this a parameter, which they call the \textbf{beta ($\beta$) factor}. The $\beta$ factor represents this dependency: The larger $\beta$ is, the more dependent is the transmission result on previous ones. The paper did not describe an updated radio transmission simulation model but it can easily be done analogous to Equation~\eqref{eq:rim} and this is one of our identified future directions for research (see Section~\ref{sec:future}).

On the other hand, great effort has been invested toward the deterministic part of the path loss $\overline{PL}$. Here, many models exist for reflecting the behaviour of the signal in various environments, such as outdoor free space, outdoor city, indoor, etc. Again, the more realistic the model is, the larger its overhead in terms of processing and memory usage. For simplicity, we omit the mean sign over the path loss in the next equations.

The simplest model is the so-called \textbf{Free Space Path Loss (FSPL)} or \textbf{Friis Equation}.

\begin{equation}
PL(d)[dB] =  -10 \cdot log\left[ \frac{G_T G_R \lambda^2}{(4\pi)^2d^2} \right]
\end{equation}

\noindent
where $\lambda$ is the signal wavelength in meters and $d$ is the distance between transmitter and receiver.

The assumption here is that the receiver and transmitter are in line-of-sight with each other and that there are no objects around to cause reflection or diffraction.
It is important to understand that this equation is not defined for $d=0$ and it only holds for larger distances, the so-called far field ($d \gg \lambda$).

Obviously, the free space path loss increases with  a factor of square of the distance. If we have already calculated or measured the path loss at some distance $d_0$, we can easily calculate the path loss at some other distance $d>d_0$ without having all parameters of the system, such as antenna gains:

\begin{equation}
P_R(d) = P_R(d_0) \left(\frac{d}{d_0}\right)^2
\end{equation}
\begin{equation}
    PL(d)[dB] = PL(d_0)[dB] + 10 \cdot 2 \cdot log\left(\frac{d}{d_0}\right)
\end{equation}

Since the expression in dB is logarithmic, this model is also called \textbf{Log Distance Path Loss}. Note that we have written $10 \cdot 2$ in the equation, instead of simply $20$. This is due to the next extension to the model, which reflects how well a signal penetrates a particular medium. Until now, we have considered a vacuum, which has a \textbf{Path Loss Exponent} of 2. Other environments, such as a grocery store, has a lower path loss exponent, typically around 1.8, while a crowded office environment can have a path loss exponent of 3.0. The generalized equation is then:

\begin{equation}
    PL(d)[dB] = PL(d_0)[dB] + 10 \cdot n \cdot log \left(\frac{d}{d_0}\right)
\end{equation}

\noindent
where $n$ is the path loss exponent.

Once the received signal strength has been calculated at the receiver's side, the propagation model can proceed with comparing it to the sensitivity threshold of the simulated radio transceiver. If the received power was below the threshold, the transmission has failed. Thus, in some way, these models can be considered to be calculating a more complex shape for the transmission area than UDG.

The models can also be applied at different levels of precision. If you apply the model to packet level, the result is either a successfully received packet or a failed packet. If you apply them on a bit level, then the higher communication layers can attempt to repair individually failed bits. Again, the trade-off is between precision and processing overhead.

The models presented here are only a few, and are quite simple. There are many more models, such as those attempting to catch the Doppler effect \cite{doppler}, indoor environments with walls~\cite{lott:2001}, and many more. However, more sophisticated ones are too expensive in terms of performance to be used for large-scale OppNets simulations and thus, out of the scope of this survey. Interested readers can turn to \cite{rappa} or \cite{karl:2005}.

Finally, \textbf{Ray Tracing} is a technique from the computer graphics domain, where a scene is rendered with rays that propagate through space and get reflected~/~diffracted~/~scattered~/~etc. In fact, it is a very precise technique that delivers extremely realistic results (as we can see from 3D rendered movies). However, it requires a very precise description of the scene itself---individual objects and their form, structure, materials, etc. Of course, this can be also done for network simulation scenarios, but requires a level of detail that we typically cannot provide. Additionally, the rendering process is very slow. Nevertheless, it has been applied to network simulation and even some optimizations have been achieved~\cite{ji01,yun:2015}.

\begin{figure*}[htbp]
\begin{center}
\includegraphics[width=\linewidth]{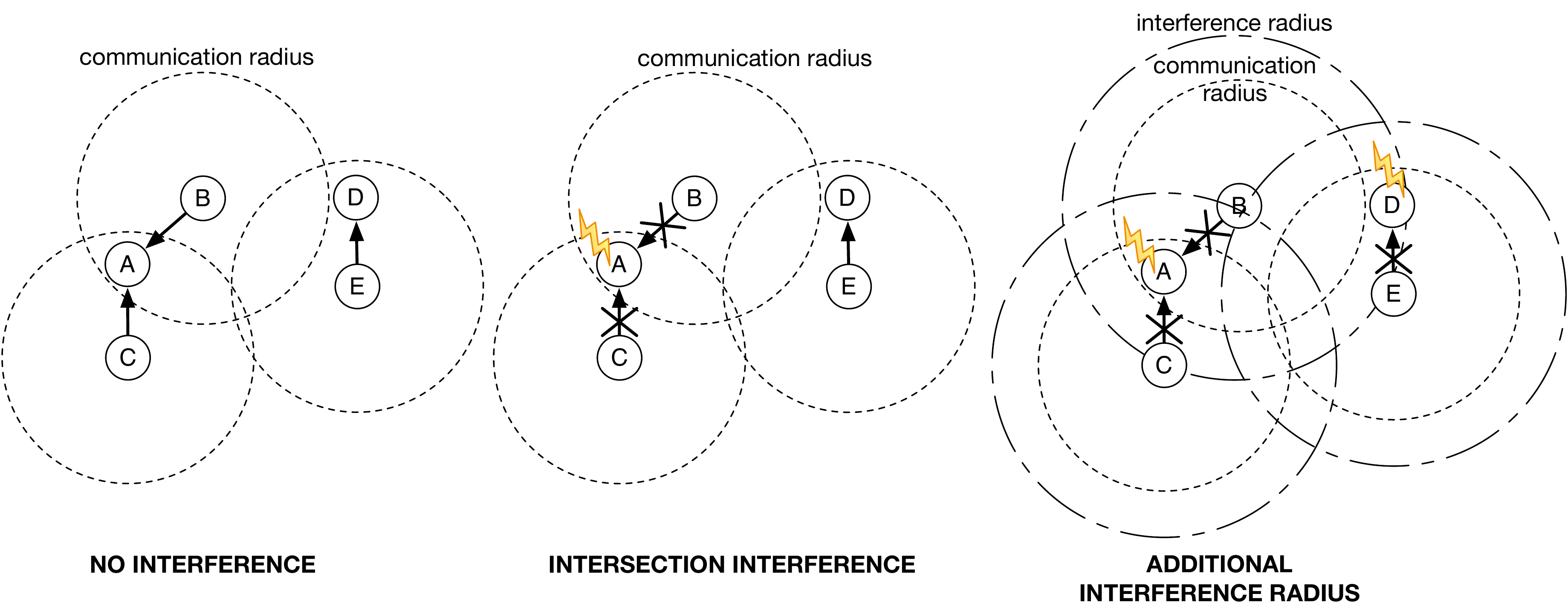}
\caption{Various interference models for the UDG radio propagation model.}
\label{fig:interUDG}
\end{center}
\end{figure*}

\subsubsection{Trace-Based Radio Propagation Models}

Another possibility for simulating radio propagation for OppNets is to gather real data from real experiments and re-run it in a simulation tool. In the context of OppNets, this can be done in two different ways: by gathering wireless link quality properties or abstracting away and gathering only the so called contact data.

\paragraph{Wireless Link Traces}
The general idea is the following: you prepare an experiment with some nodes that use the communication and link technologies relevant to your scenarios. For example, you could take 10 smartphones communicating over Bluetooth. You implement an application that simply broadcasts a packet very often (e.g.\ every 100 ms) and records the received pings from other neighbours together with their wireless quality properties, such as Received Signal Strength Indicator (RSSI) or Link Quality Indicator (LQI).

Later in the simulation, the algorithm works in reverse: When a node needs to send a packet to another node, the simulator looks up that transmission in the trace file and checks whether it was successful or not.

This way of implementing trace-based radio propagation models is mostly used for wireless sensor networks (WSNs) \cite{garg:2011wintech,Marchiori:2010}. It allows for rather good precision, but it is obviously not scalable, as it can only simulate as many nodes as the original experiment included and only in the original experimental setting. Recording such traces is also not trivial. For OppNets, the abstraction to contact times is better suited.

\paragraph{Contact times}
The idea is to abstract away from individual links and their properties and to record simply when who was connected to whom. We have already presented them in the scope of mobility traces in Subsection~\ref{sec:mobility}. For example, in the above scenario with 10 smartphones connected by Bluetooth, you will implement an application that simply checks for which other devices are around at every predefined interval (for example, every second) and records this information. Later you can process the information to identify how long a particular contact between two devices lasted. Since OppNets require exactly this type of information, such models are very well-suited and highly processing-efficient. They can also be used to analyze the properties of the individual scenarios~\cite{foerster:2012mobiopp} to abstract further away and implement a hybrid model, as described below.

\subsubsection{Hybrid Radio Propagation Models}

Analyzing trace files leads us to the next abstraction for radio propagation modelling, where a simple mathematical model is used but with parameters coming from real experiments. For example, the WSN TOSSIM simulator~\cite{levis:2003} uses a graph model, where each link is represented by an edge in the graph and assigned a bit error rate in both directions. The graph itself can easily be calculated with the help of the UDG model, as described above. The bit error rate comes from analyzing real experiments, such as in \cite{garg:2011wintech}.

This model has been widely used for simulating WSNs, but is not very popular in OppNets probably because it has a higher overhead. We discuss it further in Section~\ref{sec:future}.

\subsubsection{Performance Evaluation}

% !TEX root = ../report-ieee.tex

\begin{table*}[htp]
    \caption{A comparison between a simple UDG and a Nakagami fading model in ns-3. }
    \begin{center}
        \begin{tabular}{|c||c|c||c|c|}
        \hline
         & \multicolumn{2}{c||}{\textbf{50 nodes}} & \multicolumn{2}{c|}{\textbf{100 nodes}}\\
        \hline
         & \textbf{UDG} & \textbf{Nakagami} & \textbf{UDG} & \textbf{Nakagami}\\
        \hline
        \hline
        {Simulation Time} & 10.24~h &  16.71~h & 83.85~h  & 95~h\\
        \hline
         {Memory used} & 350 MB &  400 MB & 840 MB & 1300 MB\\
        \hline
        \hline
        {Average delivery rate} & 98\%&  99\%  &97\% & 97\%\\
        \hline
        {Average delivery delay} & 2.1~h &  1.49~h & 1.52~h & 1.22~h\\
        \hline

        \end{tabular}
    \end{center}
    \label{tab:radio}
\end{table*}%

In order to evaluate the performance of a more realistic radio propagation model against UDG, we have designed an experiment with ns-3. The results are presented in Table~\ref{tab:radio}. We use a Nakagami fading model, as readily available in ns-3, to compare with UDG. Both are configured to have an average transmission radius of 50 meters.
It can be seen that using a more sophisticated radio propagation model results in better overall network performance, especially in terms of delivery delay. The Nakagami model was also configured to a 50-meter transmission radius, but its behavior allows for communication outside this area as well. Thus, this model increases the communication opportunities between nodes, improving the general network performance. However, this comes at the cost of longer simulation times and increased RAM usage. It remains open which of the two models is actually closer to reality.

\takeaway{Sophisticated radio propagation models have a significant impact on the general network performance, but result in longer simulation durations.}

\subsection{Interference Models}

Interference is generally defined as noise in wireless communications. It can come from natural sources, like interstellar radiation or lightning, or from other devices operating in the same or close frequencies. Here we explore  inter-device interference, which has by far the larger impact on wireless propagation.

The above described radio propagation models all model individual connections, as if those were always taking place alone (no other connections running). This is especially true for all synthetic models and for wireless link traces. Contact time traces can be considered as already taking care of interference because the effect is inherently included. For example, when 10 smartphones are continuously sending and receiving ping packets, they do so at the same time and in a real environment with many other smartphones around. 

Of course, the first interference model is no model at all, which is an option that is very often used. This model is also fine to use with large-scale environments and with few connections going on at the same time because the interference impact will not be very large anyway. However, it is not a good idea for many connections and dense experiments with many nodes.

The UDG model can be extended to include interference in two different ways: intersection-based or with an additional interference radius. The difference is presented in Figure~\ref{fig:interUDG}. The left-most picture illustrates the case where three transmissions are going on and all three are successful as no interference between them is assumed. The center picture considers interference if two or more transmission radii are overlapping at one receiver (intersection model). In this case, both transmissions from nodes B and C to node A fail because they overlap at node A and cancel each other. The third option is shown at the right, where an additional interference radius is shown. It is larger than the transmission radius and assumes that a transmission might not be possible at that distance, but the signal still interferes with other connections. In this case, a transmission is considered failed if one or more transmission or interference radii overlap at a receiver.

The models presented above can also be easily used with other, non-UDG radio propagation models, such a Friis equation or RIM. There, the overlapping areas simply have shapes other than circular. When using a non-UDG model, we can also use the calculated received power at the receiver directly to decide whether a transmission is successful or not. If the difference between two transmissions is large enough, the stronger signal is received successfully. If the difference is too small, both transmissions fail~\cite{Chen:2007radio}. Note that this model is not that different from the radius-based one described above. 

For wireless trace models, we can also easily implement interference models where simultaneous transmissions to the same node get cancelled. 

\begin{figure}[htbp]
\begin{center}
\includegraphics[width = 3.5in]{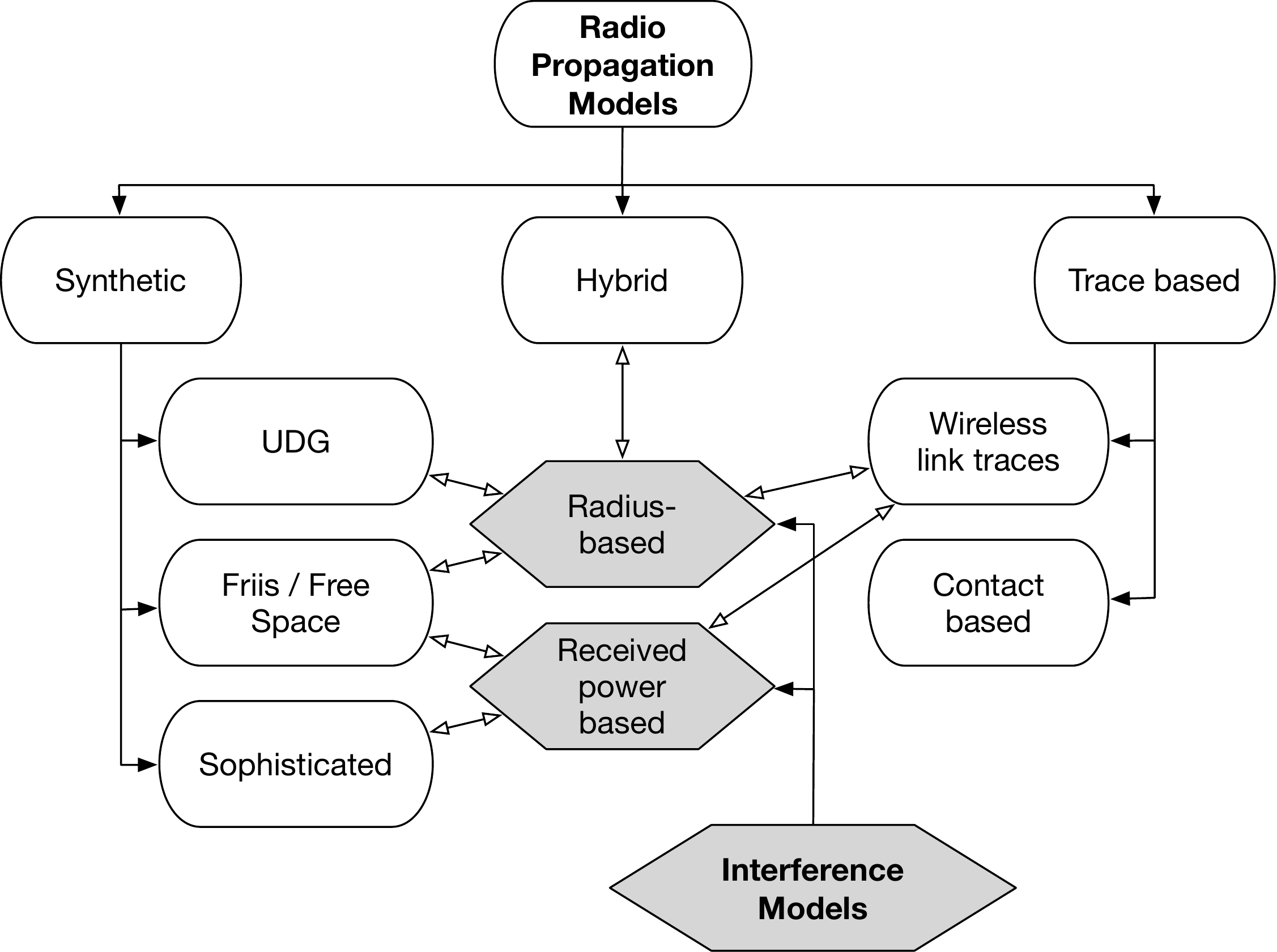}
\caption{Taxonomy of Radio Propagation and Interference Models.}
\label{fig:sum-radio}
\end{center}
\end{figure}

Figure~\ref{fig:sum-radio} presents a taxonomy of the radio propagation and interference models presented here. It is not exhaustive, but it provides a visualization of the available options and their main properties.

 %!TEX root = report-ieee.tex

\subsection{Link Technologies and Models}\label{sec:link}

The link layer as referred to in simulations and wireless networks describes a combination of the data link and the physical layer of the OSI model. The objective of the link layer is to adapt the data from the higher layers (i.e.,\ application data) for the media used (i.e.,\ wireless channel). These aspects are important when simulating OppNets since they impact the delivery rate and delays through buffer management, retransmissions, connections, etc.

The most well known link technology that can be used freely, i.e., without being bound to any operator, is WiFi technology, which is based on the IEEE~802.11 standard \cite{ieee80211-web,ieee802.11-2012} defined by the Institute of Electrical and Electronics Engineers (IEEE). Various specifications belong to this technology, the most widely known being 802.11n and 802.11ac, which are currently available on most modern handheld devices. But there exist several other specifications relevant for OppNets. For example, IEEE~802.11s is a standard specifically designed for mesh networks where the participating nodes create a layer~2 mesh network, which might also be described as a kind of \textit{MAC-relaying} infrastructure. IEEE~802.11ah is a new upcoming standard optimized for energy restricted devices such as sensor nodes and machine-to-machine communication. The focus of IEEE~802.11ah is to support a larger number of nodes  (up to several thousands) and a lower energy consumption compared to the common WiFi standard.
Relevant to OppNets is  \textit{WiFi direct}, a technology that enables WiFi devices to connect directly, allowing an easy pairing of devices for short-term data transmission without the need of an infrastructure. It basically uses the ability of modern handheld devices to become an access point. This technology was developed to overcome the practical limitations of the \textit{Ad-Hoc mode}, originally defined in the IEEE 802.11 standard with the same objectives. From the link layer point of view, WiFi direct based networks behave like classic WiFi networks.

Another widely used technology is Bluetooth, in all its variants \cite{bluetooth2016}. Especially with version 4, known as \textit{Bluetooth Low Energy} (BLE) and with the new features introduced with Bluetooth~5.0, this offers energy efficient functionality for wireless communication in constrained environments and OppNets.

The main advantage of Bluetooth and WiFi is their wide availability on a variety of end-user devices, ranging from smartphones to IoT devices. However, there are some other specialized technologies that should also be considered for OppNets since they may become relevant for some specific applications scenarios and environments, such as national park monitoring or data gathering in less densely population areas.

First of all, referring to the very active area of Wireless Sensor Networks and \textit{Smart Things}, we would like to highlight IEEE~802.15.4 \cite{802.15.4-2011}. Several higher level protocols like \textit{ZigBee}, \textit{WirelessHART} and \textit{Thread} are based on this standard. Its main advantage is the low energy consumption and optimization for low power devices. 

Currently, LoRa \cite{loraalliance-web} and Sigfox \cite{sigfox-web}, two standards that fall into the category of \textit{LPWAN} (Low-power Wide-area network), are becoming more and more widely used in the area of IoT as they claim to have a very long transmission range, up to several kilometres. The low-bandwidth offered, the possible scalability issues and the actual benefits of such long transmission ranges are still being observed and tested, but in the future they may become a good option for implementing specific OppNets-based services.

Some simulators offer realistic implementations of the link technologies described above. WiFi is well supported by OMNeT++ (using INET) and ns-3. Both can simulate Ad-Hoc and mesh networks as well as the classic, infrastructure-based ones. The new IEEE~802.11ah standard is not yet officially being supported by both simulators. IEEE~802.15.4 is also supported by OMNeT++ and ns-3; the focus of ns-3 is more on the network layer and IP-based networks (6LoWPAN) whereas OMNeT++ simulates down to the physical layer. Bluetooth is not directly supported by OMNeT++ nor by ns-3. For both simulators, there exist several community-driven projects with different qualities, in which the variety of OMNeT++ based ones is higher. Thus, implementing those link technologies is possible and typically requires only a good understanding of the standard itself.

The impossibility of completely and exactly modelling any specific real scenario in which various details must be considered, such as the presence of physical obstacles, the mobility of the nodes or the density of the nodes, leads to the use of a so-called ``ideal link layer'' for simulating high-level data propagation OppNets protocols and algorithms. Basically, most of the simulators, including those discussed in this survey, assume that if two nodes are in \textit{contact}, i.e., if their distance is less than a certain threshold, then they can exchange messages. Typically the presence of a contact is re-validated every $x$ seconds. The ONE and Adyton only have ideal link models. ns-3 has only real link technologies and the implementation of an ideal one is not trivial. OMNeT++ has the potential of offering both environments, but it requires a completely new implementation of the OppNets data propagation protocols; in this survey, we used the ideal link layer for OMNeT++. 

As we will see in Section~\ref{sec:results}, the results with ideal and real link models are very much comparable, even among different simulators. In conclusion, we can say that the use of such a conceptual link makes simulations more computationally efficient and yields results that allow alternative solutions to be compared with one another.

% !TEX root = report-ieee.tex

\subsection{User Behavior Models}\label{sec:users}

A user behavior model addresses the question: what happens after the user receives the data? Typical application models cover possibilities such as: delete the data, store the data and sometimes like~/~dislike the data. However, a real user behavior model goes beyond this and offers options like: decide to go somewhere (e.g., after reception of a concert notification for tomorrow), cancel an already scheduled movement (e.g., after receiving the weather forecast) or, very importantly, run away from danger. Other behavior models include answering a message, creating a message in response to some external event (e.g., sending a message to all friends to cancel the biking tour because of a bad weather forecast) and many more.

\begin{figure*}[!t]
    \centering
    \includegraphics[width = \textwidth]{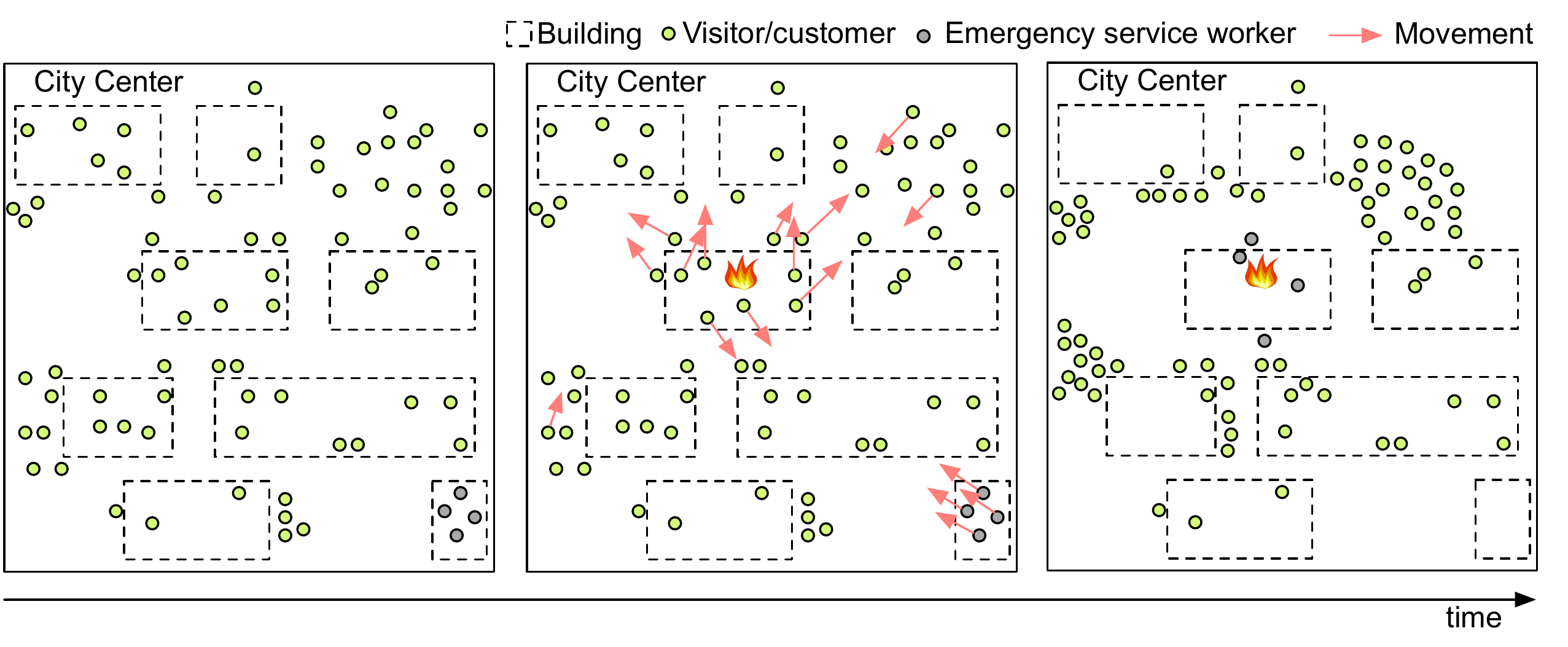}
    \caption{Implications of User Behavior: A fire in a building in the city center results in people running away from the danger zone, but gathering around it, while first responders rush toward the fire.}
    \label{fig:user_behaviour}
\end{figure*}

Why is the user behavior model important? Because it changes both the mobility and the traffic generation in the simulation. To understand this better, let us explore the example of receiving fire alarms as depicted in Figure~\ref{fig:user_behaviour}.  The initial situation is the center of a city, with people moving around. Suddenly, there is a fire alarm in one of the buildings. This alarm propagates quickly via OppNets and all users around. However, a state-of-the-art simulation does not change the behavior of people moving around---they will continue moving as if nothing has happened. The reality looks different: the first responders would run to the location of the fire to organize the evacuation, while visitors will run quickly out of the danger area and gather around it. 

There is one common property of all user behavior models: they change, at the very least, the mobility of the simulated user. As we already identified in Subsection~\ref{sec:mobility}, there are no mobility models that currently support this kind of scheduling and planning. Thus, user behavior models are also largely missing from OppNets simulations.

\takeaway{User behavior has a huge impact on application traffic and user mobility and should be modelled properly in a simulation.}

 %!TEX rolt = report-ieee.tex

\subsection{Traffic Models}\label{sec:traffic}

One of the main parameters when simulating a real-world scenario is how much data is created and circulated in a particular network.
Also in this case, the used models should follow the real-world case characteristics as closely as possible. The traffic model is considered very important in OppNets, since the creation time of a particular message is crucial for the resulting delivery delay. For example, if the user creates the message at home with no contacts to other devices, it might not be delivered until the following days when she starts moving again, thus introducing a high delivery delay.

When talking about traffic models, we need to differentiate between traffic size and traffic frequency. The former models how much data is created at once (e.g.\ always 1 kB, 1 GB or other random sizes), while the latter dictates how often data is created (once per second or week, randomly, etc.).
Additionally, it is important to note that there are destination-oriented and destination-less traffic models.

\begin{figure}[!t]
\centering
\includegraphics[width = \columnwidth]{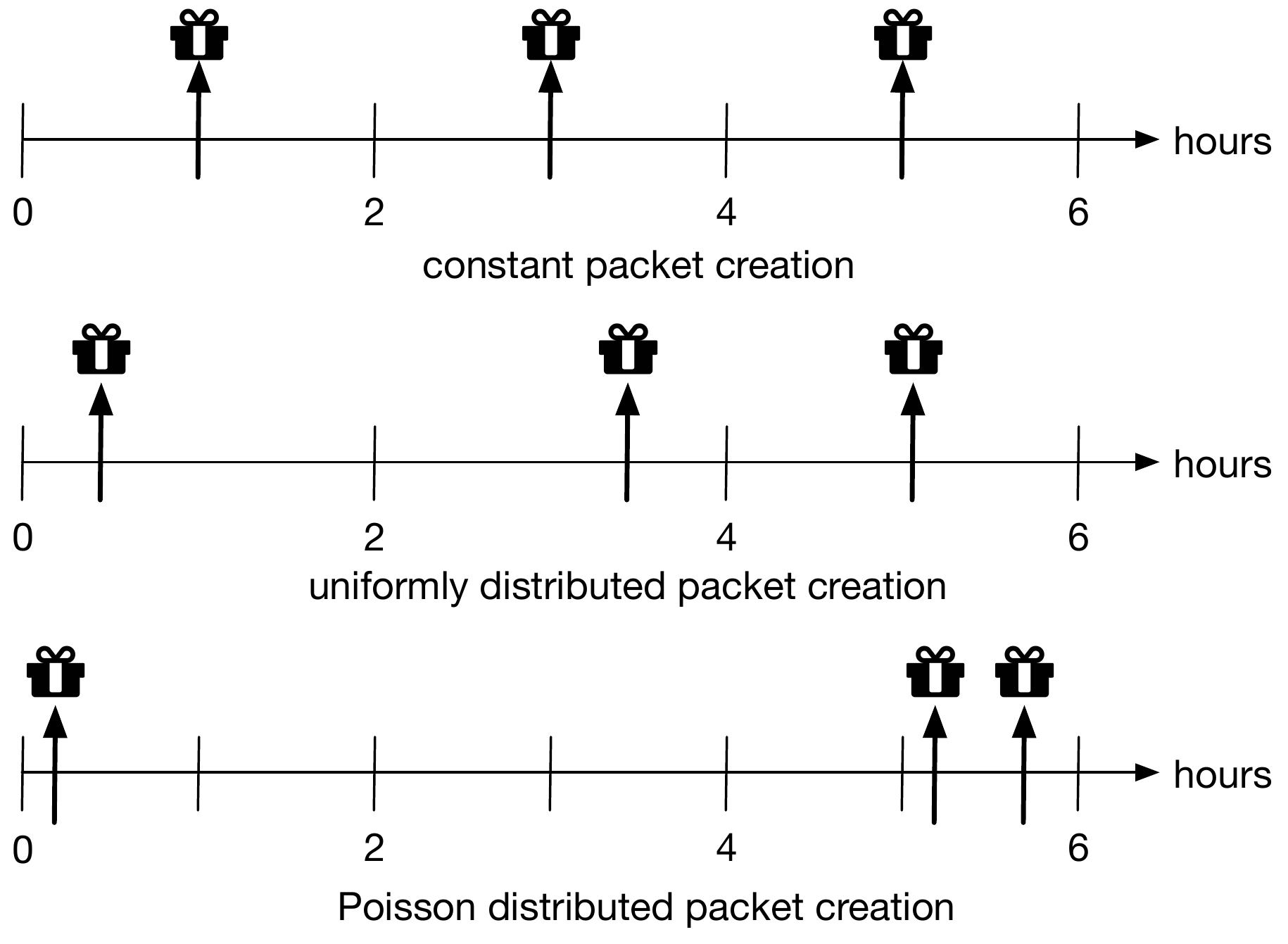}
\caption{Traffic patterns for the three most widely used traffic generators, the mean of all generators is the same.}
\label{fig:traffic}
\end{figure}

A very good overview of existing models is provided in the book of Wehrle et al.\ \cite{wehrle:2010}. We consider here only some of these models, those most relevant to OppNets, namely:
\begin{itemize}
\item \textbf{Constant (periodic) traffic:}  this is the simplest model, where a data packet is created every $x$ time interval. It is considered by many researchers as being not realistic enough but, as we will show later in this section, for very large simulations it performs equally as well as other models considered to be more realistic.
\item \textbf{Uniform traffic:} this model is also based on a constant $x$ time slot but the data packet is created any time within the time slot instead of at the beginning (or at the end) of it. The creation instant inside the slot is randomly computed using a uniform distribution.
\item \textbf{Poisson traffic:} this is probably the best known model, where the creation instant is randomly computed using a Poisson distribution. It has been shown that this distribution very closely models the traffic in user-driven network traffic, such as web browsing, phone calls, sending text messages, etc.
\end{itemize}

Figure~\ref{fig:traffic} graphically compares the three traffic patterns assuming a $2$ hour time slot. The ``Constant'' pattern generates packets periodically---at a good predefined time. The ``Uniform'' pattern generates one per 2 hour period, but the exact timing of the packet is uniformly distributed inside the interval. Finally, the ``Poisson'' pattern creates a packet at random times, thus producing sometimes very long inter-packet times and sometimes creating ``bursts'' of packets, i.e., many packets with a very short inter-packet delay.

\begin{figure}[tb]
\includegraphics[width=\linewidth]{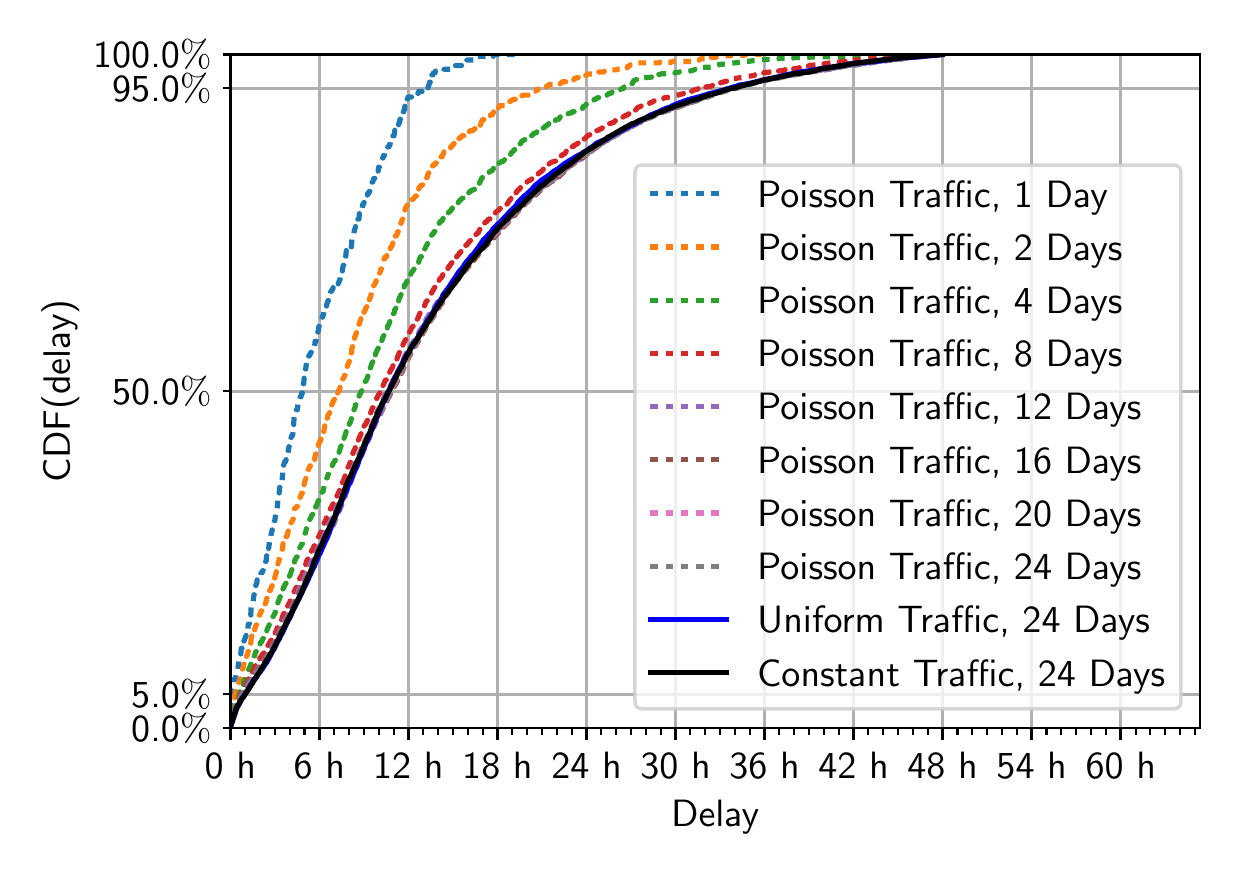}

    \caption{Performance of different traffic generators with different simulation durations. Cumulative distribution function of the delivery delay. The mean of all generators was set to 2 hours, experiments performed with OMNeT++ and 50 nodes. The packet delivery rate varies between 91\% and 97\%.}
\label{fig:traffic-experiment}
\end{figure}

As commented earlier, Poisson traffic is usually considered superior to Constant or Uniform, since it is more ``random'' and has been shown to model user-driven traffic very well. We decided to test this assumption and compared the three traffic generators using the OMNeT++ simulator and a simple opportunistic routing protocol (RRS, see Section~\ref{sec:data}). The scenario we used had 50 mobile nodes moving according to the GPS trace from San Francisco taxis, as we had already done in Subsection~\ref{sec:mobility} and will do in Section~\ref{sec:results}. The mean of all traffic generators was set to 2 hours and the cache size of every node (to buffer data) was considered to be infinite.

The resulting CDF (Cumulative Distribution Function) of the delivery delays is shown in Figure~\ref{fig:traffic-experiment}. As can be seen from the graph, when the simulation time is short (e.g., 1~day), the delivery delay of the packets varies slightly compared to the simulations of longer durations. However, when the simulation time increases, CDF curves of the delay quickly converge.
\takeaway{Constant, Uniform and Poisson traffic generators have the same impact on average performance metrics for long simulation runs.}

\begin{figure}[tb]
    \includegraphics[width=\linewidth]{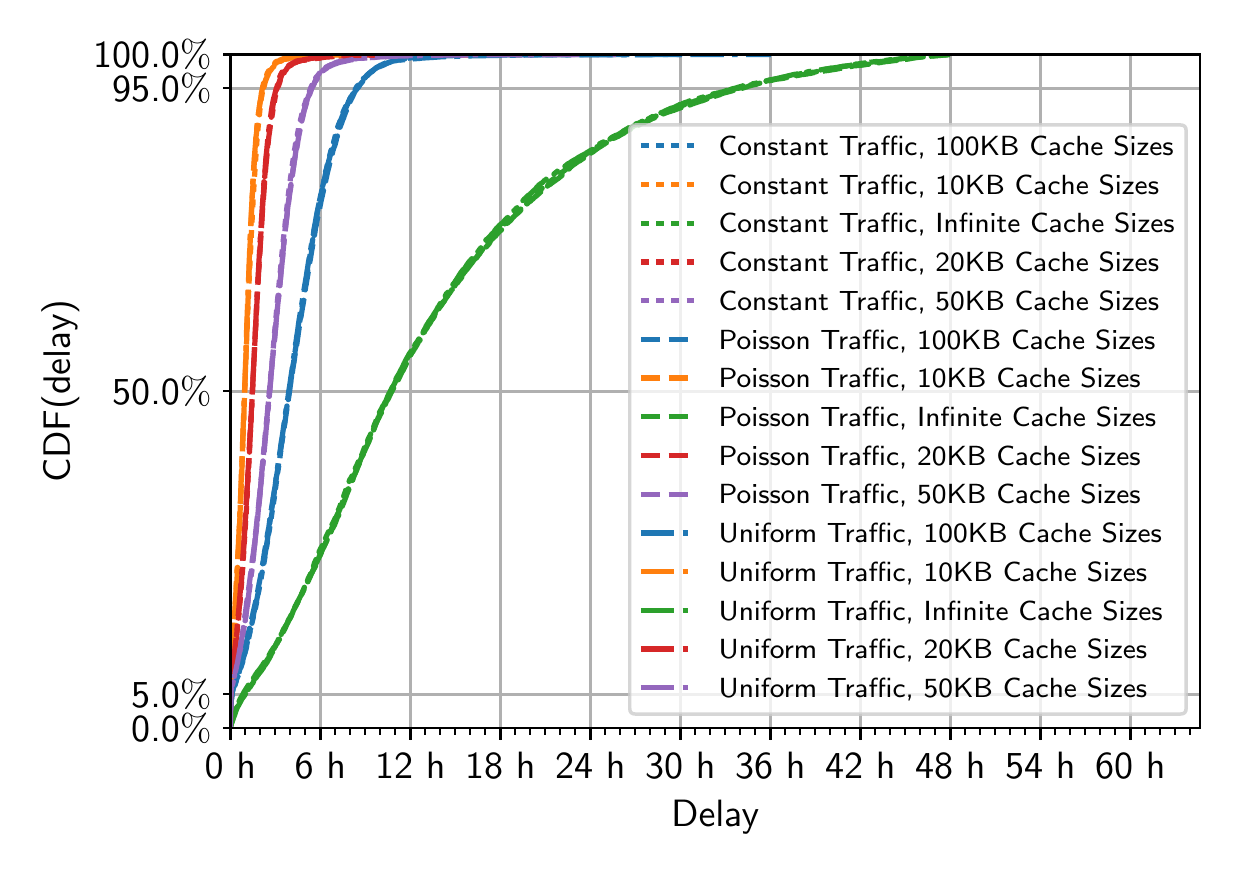}

    \caption{Performance of different traffic generators with different cache sizes. Cumulative distribution function of the transmit delay. The mean of all generators was set to 2 hours, experiments performed with OMNeT++ and 50 nodes. The packet delivery rate varies between 39\% and 92\%. It can be seen that same-coloured curves overlap, which means that there is no significant difference between the three traffic generators.}
\label{fig:traffic-experiment-buffersize}
\end{figure}

However, there is another important parameter to consider, namely cache sizes. We designed another experiment with the three traffic generators, limiting caches to different sizes (10 kB, 20 kB, and infinite). The results are shown in Figure~\ref{fig:traffic-experiment-buffersize} where it can be seen that the differences in the delays become significant when compared to the previous experiment in Figure~\ref{fig:traffic-experiment}. Moreover, when investigating  the delivery rate of these simulations, we found that the PRR changed drastically depending on the cache sizes (e.g., 10 kB caches: \textasciitilde39~\%, 20 kB caches: \textasciitilde50~\%, infinite caches (1 GB): \textasciitilde92~\%). However, different traffic generators with the same cache size showed exactly the same behavior (overlapping lines in  Figure~\ref{fig:traffic-experiment-buffersize}).

This behavior has a logical explanation, viz., OppNets have a store and forward feature, which means that when the user creates a message, this message might remain in the cache for a very long time because there are no contacts to forward it to. Thus, once the user starts moving and contacting other users, her cache is full of messages created earlier. Thus, the exact time of creation becomes less important (though it does affect the absolute delivery delay) than their number or the cache size itself.

\takeaway{Large cache sizes are crucial for achieving high delivery rates.}

The use of the previous synthetic traffic generation is the most common approach to evaluating performance, as we can simply set and regulate the average load (messages) in our simulation experiments. Nevertheless, for some evaluations (for example, considering the application usage or social aspects), we must use more complex traffic patterns, such as \textbf{trace-driven} approaches.
For example, the authors in \cite{Ristanovic12} presented a trace of 50 users using \textit{Twitter} on their smartphones with a Bluetooth-based opportunistic network.
These traces can be used to evaluate several aspects in depth, such as network usage or a device's power consumption. Nevertheless, this reproduction imposes a serious limitation on the evaluation since the reproduced traffic is fixed.

Due to these limitations, a good approach is to use \textbf{hybrid traffic}, where the traffic is synthetically generated resembling typical user application patterns. A simple approach is to generate messages with message sizes and frequencies based on known application usage statistics. For example in \cite{Herrera2016}, three message sizes and frequencies were considered: (1) a short text message (1 kB) every hour; (2) a photo (1 MB) every 18 hours; and (3) a video or high-resolution picture (10 MB) every 96 hours. These frequencies were based on the statistics of \textit{Whatsapp (Facebook, Inc.)} message usage from \cite{statista}, while sizes are approximations of the content produced by current mobile phone hardware. More statistics about the use of this kind of mobile applications can be found in \cite{Fiadino15, Zhang15}.

\subsection{Energy Consumption and Battery Models}
 %!TEX root = report-ieee.tex

All operations on end-user devices need energy. However, the exact behavior of both the energy consuming parts (processing, visualization, communication, etc.) on one hand and the batteries on the other, is very complex and exhibits non-linear stochastic properties. This complexity has resulted in the development of a number of models that characterize and emphasize different aspects of handling energy usage.
Considering the basic functionality, the available models can be classified as follows:

\begin{itemize}
    \item The \textbf{Battery Model} should ideally mimic the behavioral characteristics of real world batteries like capacity, maximum voltage and current, charging and discharging, temperature dependency, lifetime, etc.\ Simple models assume the battery functions like a bucket: a limited number of energy coins are available for usage. Once these are exhausted, the battery dies. This is usually referred to as the bucket model.
More complex models also consider the self-discharge and self-recovery effects of the battery or also take into consideration different loads and their impact on the battery lifetime. These models are covered well in the survey by Jongerden et al.\ \cite{jongerden2009battery}.
    \item The recharging of the battery is handled by \textbf{Power Generator Models}. They simulate components such as charging stations, solar panels and similar components. They are very often combined with the battery model.
        \item The \textbf{Energy Consumption} or \textbf{Energy Expenditure} models reflect the energy usage by device components and~/~or activities.  In these models, the energy consumption of the simulated devices is implemented as energy requirements of a certain state, such as idle, transmitting, receiving, processing etc., especially due to the operations of the network devices (802.15.4, Bluetooth, etc.). Depending on the granularity, the modelling of the device states can become very complex as the energy consumption of the complete device has to be taken into account. Another task of the energy consumption model is to shut down the node when the battery model signals that its capacity has been depleted, and to restart when the battery is recharged.

\end{itemize}

The complexity of these models has resulted in differing implementations in simulators.

The \textbf{ns-3} simulator supports all three models, i.e., generation, storage and consumption of energy. The generator model in ns-3 is a simple on-off generator, while there are a couple of battery models that include a model for Lithium-ion batteries according to \cite{tremblay2007generic}. Additionally, ns-3 includes an analytical battery model which implements the Rakhmatov-Vrudhula model \cite{rakhmatov2002battery}. The energy consumption is modelled in ns-3 only for the WiFi physical layer, addressing the consumption levels of the different states of WiFi.

The \textbf{OMNeT++} simulator, through the INET Framework, provides a set of models to characterize energy-related hardware properties. The framework provides generator, consumer and battery models that support a basic set of functions. Each of these models can be attached to any other module, providing it with the respective properties. Readily available models include an on-off generator (as in ns-3), an ideal battery model with infinite capacity and a residual battery model with finite capacity. Both battery models follow the bucket algorithm. The energy consumption is modelled using a two-state mechanism that supports the states of operating and sleeping, with an energy consumption parameter for both states.

The \textbf{ONE} simulator supports a model where the energy level can be set as a parameter before a simulation, together with the amount of energy required for different states such as scanning, transmitting, etc.\ of wireless interfaces. When the energy is depleted due to node activities, the node is shut down.

The \textbf{Adyton} simulator does not offer any models for energy or power consumption in its implementation.

As indicated above, modelling energy consumption and battery characteristics are complex tasks. Below, we present a simple scenario that highlights how different user behaviour patterns would influence the operations of a battery based on a specific usage and consumption pattern.

Our scenario considers the OppNets service as a secondary service on user-held devices. There are two users, Alice and Bob, who have the same OppNets chat application on their devices. Generally, the power consumption of their applications depends mainly on the device used and the activities of the applications. However, it also depends on the user herself, assuming that Alice always has all her communication interfaces switched on, as compared to Bob who switches the interfaces on only when needed. When Alice\textquotesingle s application tries to send messages, it works much faster (no power-up delay) and the overhead of using them is small as compared to Bob's. With Bob, sending a message is much more expensive due to the switch on, send, switch off cycle. As a result, the exact impact on the device battery lifetime is very hard to measure directly for individual users. Therefore, it makes more sense to consider an average overhead of OppNets services instead of lifetime or energy consumption.

\takeaway{Consider average communication and processing overhead for OppNets instead of device lifetime and energy consumed.}

% !TEX root = report-ieee.tex

\section{Propagation Protocols}
\label{sec:data}

Over the last few years, various proposals have emerged describing novel data propagation protocols for OppNets. It is not one of the aims of this survey to detail and classify all these proposals, but we will describe below the most relevant and referenced proposals, indicating the availability of OppNets simulators analyzed in this work. For a wider overview of the existing state-of-the-art in the area of opportunistic routing and forwarding, we suggest the following works: \cite{7056450} or \cite{book-spythrpicand}. Some of the more widely known protocols are presented below.

\begin{itemize}
    \item \textit{Epidemic}~\cite{Vahdat2000} is one of the most basic approaches very often used as a reference for new proposals. Its basic idea is that when two nodes meet, they first exchange a list of their data items and then synchronize them. In this way, when they divide and provided the given time was sufficient, they will have exactly the same data caches.

    \item \textit{Randomised Rumor Spreading (RRS)}~\cite{RRS} is similar to Epidemic and also floods the networks. However, when a node meets other nodes, it randomly selects one item in its cache and sends it out to all neighbors. Thus, it cannot be guaranteed that a useful item will be sent out.

    \item \textit{PRoPHET}~\cite{Lindgren2003} is a context-aware protocol that uses the history of contacts to determine the probability that a node can deliver a message to a particular destination. It can reduce the network overhead by about one order of magnitude compared to simpler approaches like Epidemic.

    \item \textit{Spray and Wait}~\cite{Spyropoulos2005}  is a simple but very effective method based on controlled replication. In this approach, the source nodes assign a maximum replication number to a message. A copy of this message is then distributed to a number of relay nodes and the process continues until one of the relay nodes meets the destination or the maximum number of replicas is reached. The idea is very efficient in controlling the overhead of individual messages. Based on this initial strategy, many other protocols were proposed that basically focus on improving the efficiency of the data delivery by using different strategies for the replication phase.

    \item \textit{BubbleRap}~\cite{Hui2011} is probably one of the first solutions that focused on using some social aspects of the users' context. They proposed the concept of \textit{community} and of \textit{centrality}. When messages are spread, nodes with higher centrality and from the same community are preferred, under the assumption that those nodes will meet the destination sooner.

\end{itemize}

Adyton supports a wide range of protocols, e.g. Epidemic, Direct Delivery, PRoPHET, Spray and Wait, SimBet \cite{Daly2007}, and BubbleRap. Furthermore, it supports many variants of Spray and Wait, e.g. Most-Mobile-First (MMF) spraying and Most-Social-First (MSF) spraying by \cite{Spyropoulos2009}, LSF Spray and Focus \cite{Spyropoulos2007}, Encounter-Based Routing \cite{Nelson2009}, Delegation Forwarding \cite{Erramilli2008a}, Coordinated Delegation Forwarding \cite{Papanikos2014}, and some others.

The ONE also offers a wide variety of protocols, e.g. Epidemic, Spray and Wait, PRoPHET, PRoPHET v2 \cite{Lindgren_Prophet2}, First Contact \cite{Jain2004}, Direct Delivery, Maxprop~\cite{Burgess} and some others.

OMNeT++ provides a number of data propagation protocols for OppNets. However, most of them are not compatible with the newer versions of OMNeT++, as explained in Subsection~\ref{sec:tools-omnet}. There, we find implementations of Epidemic, Publish-Subscribe \cite{OPPONETCACHE}, ExOR \cite{OPPONETFRAMEWORK}, and MORE \cite{OPPONETFRAMEWORK}. The OPS framework\footnote{https://github.com/ComNets-Bremen/OPS} is up-to-date and offers an RRS implementation.

Several implementations are also available for ns-3, e.g. Bundle \cite{DTN_rfc5050}, Licklider~\cite{rfc5326} and Epidemic~\cite{epidemic-ns-3-review}.

% !TEX root = report-ieee.tex

\section{Performance Metrics}
\label{sec:metrics}

Performance metrics provide the means for evaluating the performance of a given system. The type of metrics to use in evaluating a system differs from system to system. Due to the nature of OppNets, commonly used metrics for evaluating networks (e.g., throughput) have a lesser importance than certain others do. In the following, a list of widely used metrics are presented together with how they might be computed.

\subsection{Delivery Delay}

Delivery delay provides a metric of how fast a message can be delivered to an intended recipient (or a group of recipients) considering a per-node scope or a per-network scope. For example, in case of emergency messages, it is always critical to know the speed of data propagation, while in other scenarios, it is important to know how far the data is propagated depending on different node densities, mobility patterns, user preferences and so on. The timeliness of data can be evaluated as \textit{Average delay time to receive certain data per node} and \textit{Average time to propagate data through the network}. The network-wide \textit{Mean Delivery Delay} \(\delta\) is computed in Equation~\eqref{eq:deliveryDelayNetwork} where \(N\) is the number of nodes in the network.

\begin{equation}
\delta = \frac{1}{N} \sum _{i=1}^{N} {\delta_{i}}
    \label{eq:deliveryDelayNetwork}
\end{equation}

\noindent
where \(\delta_i\) refers to the \textit{Node Delivery Delay} and is computed according to Equation~\eqref{eq:deliveryDelayNode}.

\begin{equation}
\delta_{i} = \frac { 1 }{ M_{rx,i} } \sum  _{ j=1 }^{ M_{rx,i } }{ (\tau_{rx,i,j}-\tau_{tx,i,j}) }
\label{eq:deliveryDelayNode}
\end{equation}

\noindent
where \(M_{i}\) is the total number of packets (i.e., messages or data) received by the \(i^{th}\) node in the network. The transmission time \(\tau_{tx,i,j}\) is the time when the origin generates the \(j^{th}\) packet. The reception time \(\tau_{rx,i,j}\) refers to the time when the packet is received by the node \(i\).

In the case where a destination-less scenario is used, the delivery rate per node is computed according to  Equation~\eqref{eq:deliveryDelayDestLess}.

\begin{equation}
\delta_{i} = \frac { 1 }{ M } \sum  _{ j=1 }^{ M }{ (\tau_{rx,i,j}-\tau_{tx,j}) }
\label{eq:deliveryDelayDestLess}
\end{equation}

where \(M\) is the total number of messages created in this network, \(\tau_{tx,j}\) is the time the \(j^{th}\) message was generated and \(\tau_{rx,i,j}\) is the time node \(i\) received it. 

In addition to computing only mean values, it is also helpful to plot the CDF of the delays of individual nodes and all nodes in the network.

\subsection{Delivery Rate}
The delivery rate or reception ratio provides a metric of how many packets were delivered to an intended recipient before the packet was removed from the network. One main reason for the removal of a packet from the network is due to the data in the packet reaching its expiration time (time-to-live, TTL). There are other reasons as well, such as the removal from caches of the nodes due to their limited sizes and the caching policies adopted. The Delivery Rate \(\eta\) is calculated according to Equation~\eqref{eq:deliveryRate}.

\begin{equation}
\eta = \frac {M_{rx,i}}{ M_{tx,i}}
\label{eq:deliveryRate}
\end{equation}

\noindent
where $M_{tx,i}$ is the total number of packets created by all the nodes in the network with destination node~\(i\) and $M_{rx,i}$ is the number of packets received by the intended recipient node~\(i\).

In the case of destination-less scenarios, the delivery rate \(\eta\) is computed according to Equation~\eqref{eq:deliveryRateDestless}.

\begin{equation}
\eta = \frac {M_{rx,i}}{ M}
\label{eq:deliveryRateDestless}
\end{equation}

where \(M\) is again the total number of messages created in this network and $M_{rx,i}$ is the number of messages received at node \(i\).

\subsection{Overheads}

The term overhead refers to the additional activities needed to achieve an intended objective. In OppNets, the \textit{Overheads} are a measure of how much of these additional activities are required to deliver packets to the intended recipients. The following are some of the sub-metrics relevant to OppNets w.r.t.\ overhead computations.

\begin{itemize}
    \item \textit{Overhead of irrelevant data and duplicated data per node}: in OppNets, nodes also receive uninteresting data and duplicated copies. Therefore, the percentage of the number of \textit{irrelevant data received~/~forwarded} and the \textit{number of duplicated copies received} with regard to the total number of data received~/~sent gives an indication of overhead per node.
    \item \textit{Overhead of cache usage}: this shows the utilization of the cache for OppNets with regard to total memory available in the node. This can be further analyzed with regard to the types of data that are propagated. For example, after receiving the data, even if the node is not interested in the data, it might have to store and carry the data until it encounters another node to forward because others might be interested in it. This will also cause an overhead in cache usage for irrelevant data w.r.t.\ the preferences of a node.
    \item \textit{Overhead of energy consumption}: All of the above overheads also result in battery usage, e.g. for sending out messages. However, other activities of the OppNets use energy as well -- for comparing newly arrived messages with already existing ones, for sorting caches, for learning activities, etc.
\end{itemize}

Additionally, to simply compute the number of irrelevant data stored or forwarded, we could also compute the fairness of overhead among all nodes in the network. We may use the Coefficient of Variation (CoV) as a means of determining the balance of the load in the network. The \textit{Load Balancing Metric} (LBM) is computed in Equation~\eqref{eq:loadBalanceMeasure}.

\begin{equation}
    \begin{split}
        \mbox{LBM} & = CoV(f(i)) \\
        & = \frac{\sqrt{Var[f(i)]}}{E[f(i)]}
    \end{split}
    \label{eq:loadBalanceMeasure}
\end{equation}

\noindent
where \(f(i)\) is the function that returns the number of packets forwarded at a certain node $i$. Smaller values of LBM are an indication of the loads being well spread over the network, while larger values are an indication of unevenly spread loads. The same can also be done for cache overhead or, in general, for energy overhead.

% !TEX root = report-ieee.tex

\section{Comparative Study of Simulation Tools}\label{sec:results}

In this section we first describe a use case for simulating an opportunistic network that we use later for comparing the performance and output of the four different simulators we explore in this survey.

\begin{figure*}[ht]
    \centering
    \subfloat[San Francisco (Overview)]{
        \label{fig:zoom_11:a}
        \includegraphics[width=0.45\textwidth]{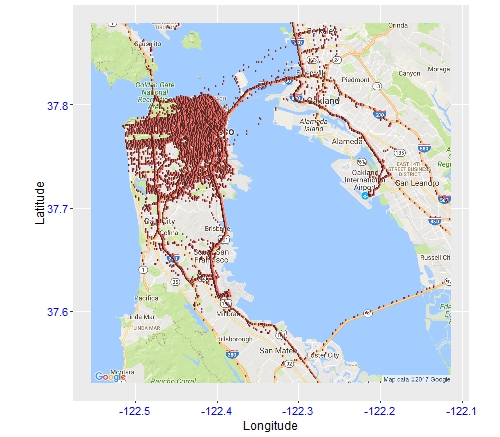}}
    \qquad
    \subfloat[San Francisco (Zoom-In)]{
        \label{fig:zoom_13:b}
        \includegraphics[width=0.451\textwidth]{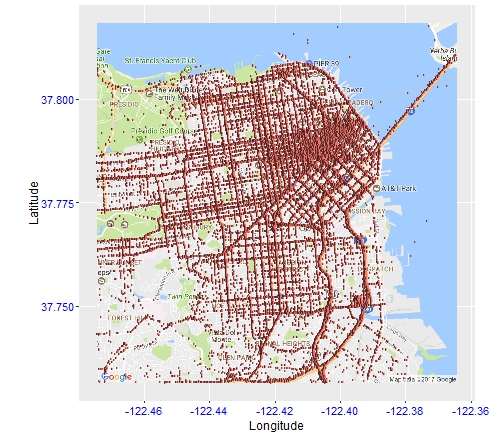}}
    \caption{Vehicular GPS trace sample (San Francisco taxis-cabs).}
    \label{fig:sf_trace}
\end{figure*}

\subsection{Use Case Scenario}

% !TEX root = ../report-ieee.tex

\begin{table*}[htp]
\renewcommand{\arraystretch}{1.1}

\begin{center}
\begin{scriptsize}
\caption{Summary of the used simulation models for all simulators and their main parameters.}
\label{tab:scenario}

\begin{tabular}{|c |c | c | c | c |}

\hline
\textbf{Model}  &\textbf{Adyton} & \textbf{ONE} & \textbf{OMNeT++} & \textbf{ns-3}\\
\hline
Mobility & SFO trace, 24 days,  & SFO trace, 24 days,  & SFO trace, 24 days,  & SFO trace, 24 days, \\
 & contact trace & 1 sec position update & 1 sec position update & on position update request (1 sec)\\
\hline
Radio propagation &  UDG, 50 m & UDG, 50 m & UDG, 50 m &UDG, 50 m\\
\hline
Interference & none & none &none &none\\
\hline
Link technology & direct contact, 1 sec scan  & direct contact, 1 sec scan & direct contact, 1 sec scan & WiFi \\
\hline
Data propagation & Epidemic& Epidemic&RRS &Epidemic\\
\hline
Application & Single destination & Single destination& Single destination&Single destination\\
\hline
User behavior & none& none& none& none\\
\hline
Traffic & Exponential, 1 pkt / 2 hours & Poisson, 1 pkt / 2 hours& Poisson, 1 pkt / 2 hours&Poisson, 1 pkt / 2 hours\\
\hline
Power consumption & none & none& none&none\\
\hline
Battery & none& none&none &none\\

\hline
\end{tabular}
\end{scriptsize}
\end{center}
\end{table*}%

We used a popular OppNets scenario in which people are moving through a city and exchanging short messages about events in the city. A typical message might be: "Coffee sampling event at Coffee Corner, 23.03.2017, 2-6 pm". We varied the number of nodes to test the scalability of the tools. In terms of simulation models, we used the following configuration (summarised also in Table~\ref{tab:scenario}):

\subsubsection{Mobility Model} For our comparison we used a trace-based model (see Subsection~\ref{sec:mobility}). Here also comes the first challenge for implementing our scenario. In order to evaluate the different simulators, we required a large trace with many nodes over the course of many days. We evaluated different traces, and the best option was the mobility trace of taxi cabs in San Francisco, USA \cite{SFO_Cabs2009} (also referred to as SFO trace), due to its high resolution, number of nodes and duration. This dataset contains GPS coordinates of approximately 500 taxis collected over 24 days in the San Francisco Bay Area, as shown also in Figure~\ref{fig:sf_trace}. 
This trace has been processed with the BonnMotion tool~\cite{bonnmotion} in order to generate the appropriate trace format required for each evaluated simulator. However, it must be noted that each simulator \textit{uses} the trace in different ways. For example, Adyton pre-computes the contacts between individual nodes and uses this contact trace for its simulations. OMNeT++, ns-3 and the ONE have different parameters as to how often they re-compute the current position (move on a straight line between two trace points). OMNeT++ and the ONE update the position according to a parameter, while ns-3 updates the position every time a higher layer requests it. In our case, this is approximately every second.

\subsubsection{Radio Propagation and Interference Model} We used Unit Disk Graph (UDG) with a communication radius of 50 meters. We do not consider interference.

\subsubsection{Link Technology} For scalability reasons and with the takeaway from Subsection~\ref{sec:link} in mind, we do NOT use any communication protocol. However, message transmission depends on the bandwidth (fixed to 2.1 Mbps) and the contact duration calculated from the traces, in order to obtain a more realistic evaluation. This simple model checks who is around every second and communicates directly with these neighbors while the contact is still there. This was not possible for ns-3, where we use the complete IP stack and a WiFi implementation. Furthermore, Adyton does not have any link technology as it pre-computes contact traces from GPS traces.

\subsubsection{Data Propagation} Epidemic~\cite{Vahdat2000} is used for Adyton and the ONE. In OMNeT++, the RRS implementation is the only one available, which we use in this evaluation. 
For ns-3, we use an implementation of Epidemic routing according to \cite{alenazi2015epidemic}, which is based on WiFi Ad-Hoc and IP traffic. It is implemented as an IPv4 routing protocol.

\subsubsection{User Behavior Model} The assumption is that the user simply reads the message.

\subsubsection{Traffic Model} The users produce the messages with a Poisson traffic generator with a mean of one message per two hours. The size of the message is 1 KB. There is a single, random destination for each message.

\subsubsection{Power Consumption and Battery Model} We do not consider these in our scenario.

In terms of  metrics, we compared the output of the simulators in terms of delivery latency (delay) and delivery rate at all nodes after simulating the complete 24 days of traces. We started with a baseline scenario of 50 nodes and we varied the number of users between 50 and the maximum of 536 nodes, which corresponds to the total number of available traces. Furthermore, we compared the memory used and the real (i.e., wall clock) time needed to simulate the complete 24 days of traces.

All experiments have been performed on virtualized Linux servers with 8~processor cores (Intel Xeon, Sandy Bridge) with 48~GB RAM. The operating system was Ubuntu Yakkety 16.10 using a 4.8 standard kernel. For the virtualization, KVM (versions~2.1) has been used.

In general, the parametrization of ONE and Adyton was straightforward and relatively easy to accomplish. These two simulators are clearly targeted towards OppNets and provide exactly the parameters we needed, without options for link technologies, radio propagation models, power consumption, etc. OMNeT++ was also not difficult to setup and parametrize, but we are biased in this case because of our year-long experience with it. The setup of ns-3 proved to be the most difficult of all because of the higher complexity (IP based simulation) and the resulting higher number of parameters that influence the overall simulation results.

\subsection{Scalability}

\begin{figure}[t]
    \begin{center}
        \includegraphics[width=0.45\textwidth]{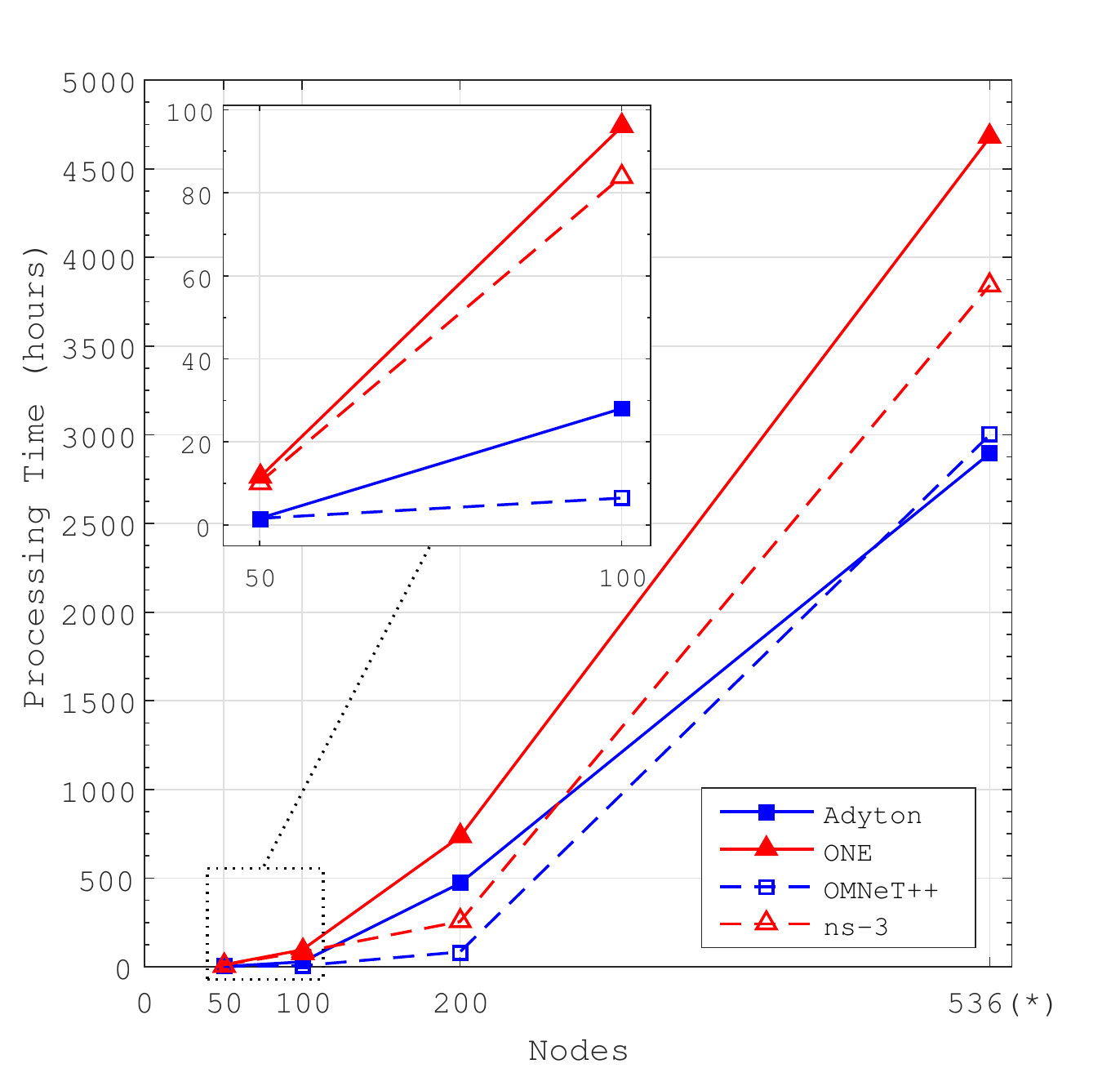}
        \caption{Simulation time (i.e., wall clock time) required for the four simulators for 50-536 nodes.\newline {\small $*$ Values are extrapolated}}
    \label{fig:time-all}
    \end{center}
\end{figure}

\begin{figure}[t]
    \begin{center}
        \includegraphics[width=0.45\textwidth]{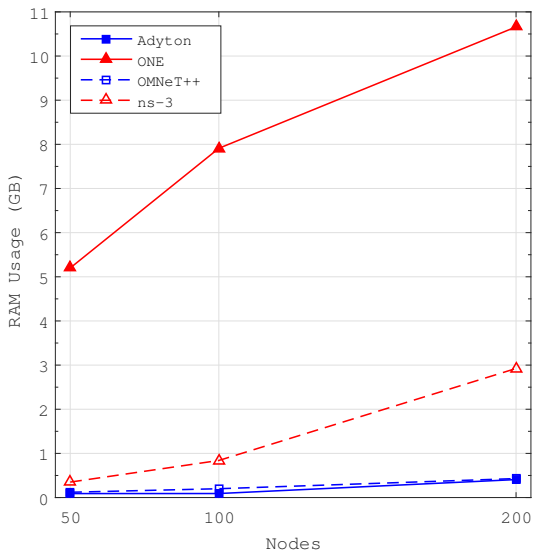}
        \caption{RAM used for the four simulators for 50-200 nodes.}
        \label{fig:ram-all}
    \end{center}
\end{figure}

We start the comparison between the simulators in terms of their own scalability and performance. Figure~\ref{fig:time-all} shows the simulation time depending on the number of simulated nodes.
Note that the values for the simulations with 536~nodes are extrapolated as none of the four simulations have completed even after 4 months. The extrapolation has been done after perusing the logs generated up to now.
There is a clear message in this graph: the currently available tools are not scalable--536 nodes is not a lot and 24 days is realistic.

There is also a clear trend to be seen: simpler models and using C++ as the programming language provide a significant advantage. As such, the ONE is the slowest (the only one written in Java), followed by ns-3, which has a complex link technology model, then comes Adyton and OMNeT++ is the fastest. Note that OMNeT++ has been optimized for performance over many years, while Adyton is a newcomer. Thus, we might expect that Adyton's performance will improve dramatically soon. Regarding the memory used, we can see in Figure~\ref{fig:ram-all} that the ONE uses more than 5 times more RAM than the others. It is followed by ns-3 because of the IP stack simulation. However, all of the simulators stay within reasonable limits in their RAM usage, even for larger simulations. At the same time, it is probably not advisable to use a simple desktop machine for such large-scale simulations.

Additionally, we explored the size of the trace files for each simulator and summarized the results in Table~\ref{tab:trace_sizes}. As can be seen, the ONE uses huge trace files, while the files for ns-3 and OMNeT++ are comparable to the original file. It can also be observed that using contact traces instead of GPS traces plus radio propagation models saves a lot of memory (Adyton). We must point out here that Adyton in fact \textit{loads the complete contact trace} into RAM before starting the simulation and still, its memory usage is one of the lowest (Figure~\ref{fig:ram-all}).

\begin{table}[htp]
\renewcommand{\arraystretch}{1.1}

\begin{center}
\caption{Trace file sizes for all simulators compared with the size of the original trace.}
\label{tab:trace_sizes}
\begin{scriptsize}

\begin{tabular}{|c |c | c | c |}

\hline
\textbf{Trace file} &\textbf{Size, 50 nodes} &\textbf{Size, 200 nodes} &\textbf{Size, 536 nodes} \\
\hline
 \textbf{Original trace} &87 MB&323 MB&876 MB\\
 \hline
 \textbf{Adyton} & 0.88 MB & 16.03 MB & 117.6 MB\\
 \hline
 \textbf{ONE} & 4,800 MB &20,000 MB & 52,000 MB\\
 \hline
  \textbf{OMNeT++} & 49 MB&183 MB& 494 MB\\
  \hline
  \textbf{ns-3} & 99 MB & 371 MB& 1009 MB\\
\hline
\end{tabular}
\end{scriptsize}
\end{center}
\end{table}%

\takeaway{Current tools are not scalable when handling the dimensions that OppNets require (thousands of nodes).}

\subsection{Performance Metrics}

\begin{figure}[t]
    \begin{center}
\subfloat[] {
        \includegraphics[width=0.45\textwidth]{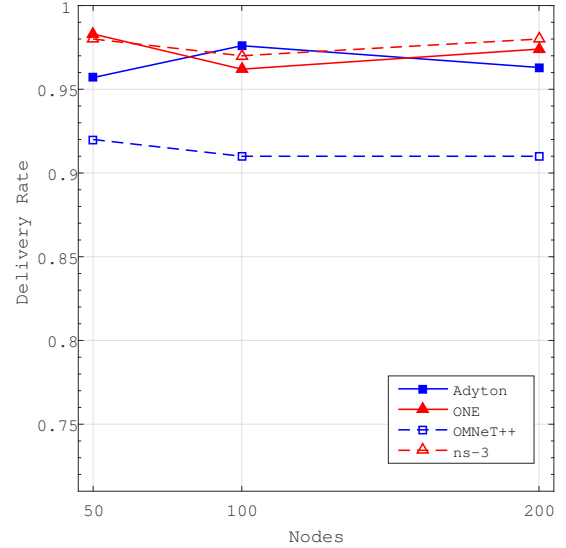}
      \label{fig:delivery-all}}

\subfloat[] {
        \includegraphics[width=0.45\textwidth]{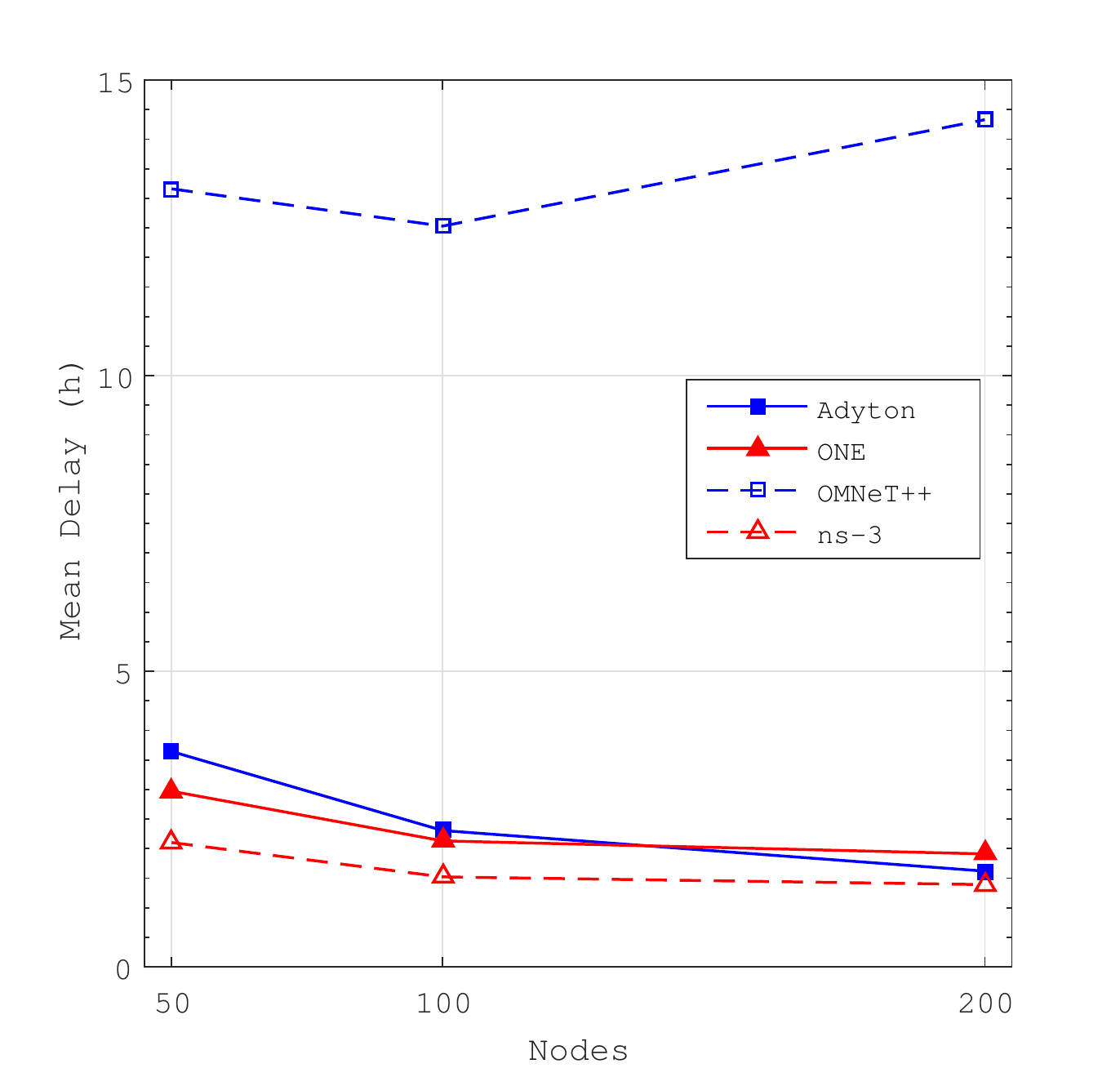}
        \label{fig:delay-all}}
    \end{center}
\caption{Delivery Rate (a) and Mean Delay (b) for the four simulators for 50-200 nodes.}
\label{fig:metrics-all}
\end{figure}

Let us now turn to the OppNets evaluation results, i.e., the obtained metrics from the different simulators.
Figure~\ref{fig:metrics-all} presents the delivery rate and mean delay results for all scenarios and all simulators. The outlier is clearly OMNeT++ and the reason is the slightly different data propagation model (RRS instead of Epidemic for all others). This is also the reason that the delivery delay for OMNeT++ is much higher.

\takeaway{Slight differences in the data propagation model result in large performance differences. Be aware of the exact implementation of a particular protocol!}

\subsection{Impact of Trace Resolution}

We could potentially reduce the resolution of our mobility traces to overcome the performance problems of the simulators. We decided to test this hypothesis. The original traces have a resolution of one second. Thus, we generated a new trace with a resolution of five seconds, comparing the performance of a simulation with both traces. The results are summarized in Table~\ref{tab:resolution}. As can be seen, reducing the resolution of the trace file from one second to five seconds in fact reduces the simulation time significantly, especially for the ONE. However, the results have also changed: in particular, the delivery delay increases very significantly. For OMNeT++, the delivery rate also decreased. The reason of this behavior is clear: for greater resolutions some of the possible contacts are missed, so the opportunity to transmit a message is lost, thus reducing the overall performance.

\takeaway{Results from the same trace file but with different resolutions are not comparable with each other.}

% !TEX root = ../report-ieee.tex

\begin{table}[htp]
\renewcommand{\arraystretch}{1.1}

\begin{center}
\begin{scriptsize}
\caption{Comparison between 1 and 5 seconds resolution of the trace files.}
\label{tab:resolution}
\begin{tabular}{|c |c | c | c | c |}

\hline
\textbf{Simulator}  &\textbf{ONE} & \textbf{ONE} & \textbf{OMNeT++} & \textbf{OMNeT++}\\
\hline
 \textbf{Resolution} &\textbf{1 sec } & \textbf{5 sec } & \textbf{1 sec } & \textbf{5 sec }\\
\hline
Delivery rate & 98\% & 97\%& 92\%& 79\%\\
\hline
Delivery delay & 2.9~h& 6.5~h& 13.16~h& 19.01~h\\
\hline
RAM & 5.2 GB& 5.14 GB& 0.127 GB& 0.124 GB\\
\hline
Simulation time &11.43~h &2.03~h & 1.81~h& 0.7~h\\

\hline

\end{tabular}
\end{scriptsize}
\end{center}

\end{table}%

\subsection{Who is the fairest of them all?}

It is not easy to answer the question of which simulator is the best. Table~\ref{tab:simulators} offers a high-level comparison between the simulators, their readily available models and their user friendliness. This table can be used as an inital overview to compare them in terms of desired requirements, programming skills and application scenarios. However, what is probably more important, especially for newcomers in the area of OppNets, is whether you can get proper help. Thus, if there is already a simulator in use in your community~/~research group, even if used for other types of network simulations, you should stick to that simulator. None of those presented here is perfect for any of the relevant applications anyhow.

% !TEX root = ../report-ieee.tex

\begin{table*}[hbt]
    \begin{center}
    \caption{High-level comparison between OppNets simulators.}
    \label{tab:simulators}
        \begin{tabular}{|c||c|c|c|c|}

        \hline
        \textbf{Tool} & \textbf{Adyton} & \textbf{ONE} & \textbf{OMNeT++} & \textbf{ns-3} \\
        \hline
            \textbf{Platforms} & Linux& Java (JDK 6+)& Win, Linux, Mac & Linux, Mac, FreeBSD\\
        \hline
        \textbf{Programming language} & C++ & Java & C++, NED & C++, Python\\
        \hline
        \textbf{Parallelizable} &- &- &+ & o \\
        \hline
        \textbf{BSD / Linux API compatibility} & - & - & - & + (DCE framework) \\
        \hline
               \hline
        \textbf{Documentation} & + & + & ++ & + \\
        \hline
        \textbf{Mailing lists and tutorials} & o  & ++ & ++ & + \\
        \hline
        \textbf{User interface} & - & + & ++ & - \\
        \hline
        \hline
        \textbf{Mobility models} & o & + & ++ & o \\
        \hline
        \textbf{Radio propagation models} & - & o & + & + \\
        \hline
        \textbf{Interference models} & - & o & + & + \\
        \hline
        \textbf{Link technologies} &-  &-  & + & +\\
        \hline
        \textbf{OppNets data propagation models} & ++ & ++ & o & o\\
        \hline
        \textbf{User behavior models} & - & - & - & - \\
        \hline
        \textbf{Traffic models} & ++ & ++ & ++ & ++ \\
        \hline
        \textbf{Energy consumption models} & -  & o & + &  +\\
        \hline
        \textbf{Battery models} & -& o& +& ++\\
        \hline
        \hline
        \textbf{Scalability} & +& -& +& o\\
        \hline
            \multicolumn{5}{c}{\small - no support, o partial support, + adequate support, ++ well supported}\\
        \end{tabular}
    \end{center}
\end{table*}%

If there are no simulators already in use, then you should consider joining the development of some simulator. The most promising ones are currently Adyton and OMNeT++. While Adyton is richer in models and has very promising performance, OMNeT++ is more sophisticated and mature, with very good support, documentation and user interface. The one to choose probably depends on your programming skills and preferences. What we explicitly do not recommend is to start yet another new home-brewed simulator. There are enough choices and enough opportunities to shape the development of existing simulators.

\takeaway{Prefer a simulator with which you or your group has already experience. Publish all new protocols and algorithms you develop open-source.}

The next section also offers a list of suggestions to follow when conducting OppNets performance analysis with any of the presented simulators.

% !TEX root = report-ieee.tex

\section{Best Practices}\label{sec:best}

Throughout the text of this survey, we have provided readers with take-away messages summarizing the knowledge gained from the material presented, and we have made suggestions on which of the existing models to use and when. In the next section, we will identify what is still missing in existing models and suggest some possible concrete ideas on how these deficiencies might be addressed.

In this section we shall summarize some of this survey's further findings that we did not state previously.

\begin{itemize}
\item Mobility models are the most important driver of performance in OppNets. \textbf{Carefully select an appropriate mobility model considering the specific use case.} Preferably select scalable, sophisticated hybrid models.
\item Radio propagation, interference, battery and energy consumption models are not required for simulations in OppNets. \textbf{It is better to evaluate these factors in testbeds or to outsource the complex computations to a pre-step, as described in Subsection~\ref{sec:contactBasedMobility}.}
\item Using models for link technologies introduces too much overhead in the simulator and also affects the performance metrics of the OppNets. \textbf{Unless they are important for the evaluation per-se, it is worthwhile to use a simplified link technology.} 
\item Simulating use cases in OppNets (similar to other environments) is a game of mixing-and-matching the appropriate models and setting their parameters to suit the given use case. \textbf{Meticulously document the models and the parameters you use for later comparison and reasoning.}
\end{itemize}

Our last finding is the most important one. To this end, we are providing a sample experiment journal of one of our experiments (Table~\ref{tab:mobility-results}) in Appendix~\ref{app:log}. We have used such journals to document all of the presented experiments, which makes it so anyone can easily reproduce our results and identify strong or weak points in the setup. The logs and a blank template are available online\footnote{https://github.com/ComNets-Bremen/OppNets-Survey-Material}.
If always provided, these logs can significantly increase the readability of new results and will lead to much more productive discussions among researchers and faster progress in the research area.

% !TEX root = report-ieee.tex

\section{Future Directions}\label{sec:future}

Section \ref{sec:results} provided a description of which models are already available in which simulator. It is hard to come to a conclusion as to which simulator is best to use since none of them has the complete spectrum of models, the desired performance and feature support level. It is left to the readers to decide which of these tools to select for their work and to try to compensate for the missing bits.

In the following paragraphs, we will sketch out some concrete ideas about further developments in simulation models, irrespective of the tool being used. These ideas arose during the writing of this survey and can be used by researchers to improve the current simulation-based performance evaluations of OppNets.

\subsection{User Behavior Models}
Typically, OppNets deliver messages to users and users react to these messages. Such user behavior models, as already discussed in Subsection~\ref{sec:users}, do not practically exist for now. These models should include two basic actions that could be taken by the user:

\begin{itemize}
\item Move as a reaction to a message -- either move immediately or schedule a move in the future. Of course, the mobility model needs to be able to react to such an action. Such a user behaviour model is needed for applications delivering this type of information as well: event announcements, traffic information, danger alerts, etc.
\item Create one or several messages as a response to a message---this is typical behavior of a user who just received a message from a friend. Such  a model is important for all applications where users create messages (and not machines).
\end{itemize}

The second action is very simple to implement and only requires trivial changes to the traffic models. The first one, however, is more important as it directly influences the mobility of the user, which we have identified to be the driving force of OppNets. The model itself needs to accommodate some random but realistic behavior of the user. For example, if Alice receives 10 event announcements per day, we cannot expect her to go to all these events. Somehow the selection of the event should be driven by Alice's preferences (which might be hidden from the OppNets system), but may also be influenced by the behavior of her friends (are the friends attending too?) or some purely random selection.

\subsection{Reactive Mobility Models}

In order to support the user behavior model described above, we need a reactive mobility model capable of scheduling a move in the future or moving somewhere immediately. Any hybrid mobility model, as described in Subsection~\ref{sec:mobility}, can be changed to accommodate such behavior. What is needed is an interface between the user behavior model and its mobility model. However, mobility models based on real movement patterns (i.e., traces) are not able to do this due to their dependence on the position coordinates and the corresponding times listed in the trace.

\subsection{Realistic and Scalable Mobility Models}

In general, real mobility models (i.e., trace based) have many disadvantages, mainly related to flexibility, scalability and trace generation. Even if these mobility models are very realistic and very easy to use, OppNets performance analysis should focus on developing more sophisticated hybrid models. Here we sketch out an idea of how to combine the advantages of real and hybrid mobility models:

\begin{enumerate}
\item Extract Points of Interests (PoIs) from real traces. Geographical information systems can be used very efficiently to identify good PoIs and their contextual information~\cite{Tran:2013:AIP:2534303.2534304}.
\item Break the trace into segments, where each segment is a real movement between two PoIs; each segment will become a ``mini-trace''.
\item Build a graph where nodes are PoIs and links are segments. There might be more than one link between two nodes, as the movement between one pair of PoIs might have occurred several times and with various means of transportation or speeds.
\item For each node (PoI), extract statistics for the pause time from the original trace and attach this information to the respective nodes.
\item Define a mobility model, which starts at one PoI, pauses there accordingly, then moves on to the next PoI by following one of the available segments (mini-traces).
\end{enumerate}

This approach has two very important and novel advantages. First, different traces can be merged together incrementally over time. This will give all researchers a very large database for performing very large, realistic mobility studies. Second, the exact algorithm of traversing the graph can be changed easily and can accommodate ideas and research from current hybrid mobility models, e.g.,  moving with friends or following a day-night schedule, etc.

The performance of such a model is expected to be better than real mobility models, but worse than state-of-the-art hybrid models. This is due to the fact that the simulated user does not move on a straight line between two PoIs, but rather on a mini-trace. Therefore, smart multithread-based model implementations will be required to address issues such as efficient operation of the models, e.g.\ to load the new segment data in a separate thread while the node is pausing in the current PoI.

\subsection{Contact-based Mobility and Communication}\label{sec:contactBasedMobility}

Contact-based traces are very efficient in terms of performance, as our results show with Adyton. In fact, they combine the impact of mobility, radio propagation and link technology models into one input trace. Thus, the overhead of computation during simulation runtime is minimized significantly. However, the main question is: is the resulting behavior realistic?

Currently, there are two ways of obtaining contact traces: experimentally and model-based. Obtaining them experimentally sounds like a very good idea, though reality proves to be different, as we know from our own studies~\cite{foerster:2012mobiopp}. In such experiments, nodes regularly broadcast beacons and log beacons from other nodes. When looking into the logged beacons from a pair of nodes, we most often observe highly asymmetric behavior: node A has logged many more beacons from node B, or vice versa. A post-processing step is needed to decide when the contact actually occurred. In fact, we do not know whether and for how long the nodes would have been able to actually exchange data. For example, link setup time and re-connect time are not taken into account at all. Thus, we need more sophisticated experiments, in which the length of contact is obtained by taking into account real data communications over appropriate link technologies meaningful for the use case (i.e., Bluetooth LE instead of WiFi).

The second option for obtaining contact traces by models is perhaps more practical to achieve. Here, we would need a sophisticated simulation environment with good mobility (for example, the idea above), radio propagation, interference and link technology models. This simulation can produce contact traces similar in quality to those obtained experimentally. These contact traces can be then used for simulating large scale OppNets scenarios.

An open challenge is how to combine user behavior models with contact-based traces. One possible but complex option is to pre-compute contact data ``just in time'' in a separate thread of the simulator. Thus, if the mobility of a particular user changes, the contact data can be re-computed on the fly.

\subsection{Radio and Interference Models}

Purely synthetic radio propagation models are highly complex, not just in terms of performance, but also in terms of correct usage and parametrization; this hides a highly probable source of error. On the other hand, hybrid models that use statistical trace data are much more reliable and simple to use. Many research efforts have been put into such models for wireless sensor networks~\cite{levis:2003}, but not for OppNets. This is a clear gap to be closed soon, especially considering the above idea of simulating realistic contact traces.

The main real-world properties to be considered in such a model are:
\begin{itemize}
\item Asymmetric links, mainly due to non-circular radio propagation patterns.
\item Bursty links, where the success of the next transmission depends on the history of transmissions. This has been observed and quantified with the beta-factor in~\cite{levis07beta}.
\end{itemize}

Unfortunately, there is currently no synthetic model available that takes into account all these properties together.

\subsection{Credible and Realistic Traffic Models}

We have shown in this survey that the traffic models used do not significantly affect the results of large simulations (Subsection~\ref{sec:traffic}). We have also identified that cache sizes are crucial to increasing the delivery rate in the network. At the same time, there is a mutual impact between the size of individual data messages and real contact length (considering the link technology used as well). This connection needs to be explored.

Furthermore, the correct data size selection process seems rather random in state-of-the-art research. For example, many applications consider the exchange of large video files---but the question is: what realistic OppNets service would exchange videos? Short messages, such as text messages or small images are much more realistic and can be better served with OppNets. The exact parameters need to be extracted from real services and applications.

Moreover, even though we have shown that constant, uniform and exponential traffic patterns deliver comparable results, a realistic traffic model is still lacking. For example, an exponential traffic pattern still generates traffic during the night---but, this is very rare behavior of real users. Such models are still lacking and their impact on the performance of OppNets needs to be explored.

\subsection{Scalability}

All of the above described ideas also seek an increase in the scalability of OppNets simulations. However, while they will most probably increase the general performance, they will not conceptually allow for very much larger scale simulations, e.g., millions of nodes. Thus, more far-reaching new concepts are required.

One novel idea has been presented in \cite{fleo:2015}, where a packet-based simulation has been exchanged with a flow-based simulation. This saves a significant amount of discrete events, as only the start, the end and the status changes of a communication session between two nodes is modelled. It could be interesting to transfer this idea to OppNets simulations.

Furthermore, mathematical models are also a good alternative, especially for very large networks. We have already briefly explored the current state-of-the-art in Section~\ref{sec:perf}.

\subsection{Metrics}

Current metrics used in OppNets performance analysis are rather simple and assume that the user is willing to receive either everything (irrespective of how old or relevant the data is to her) or to receive the messages destined to her only. However, especially in OppNets applications, we often observe that people would like to receive ``relevant'' data. The question, in any case, is how to define relevance: is it the receipt of timely data or of popular data?

Generally speaking, it depends on the data itself. Expired data, such as a finished event or outdated traffic information, is not useful at all unless it was so popular that you would like to read about it later. For example, the fact that there was a traffic jam somewhere in the city yesterday is probably irrelevant. However, the news about a large demonstration in the city yesterday might be of great interest to many people.

One possibility for evaluating the performance of a particular data propagation algorithm is to allow users to evaluate data messages offline. For example, we can assume that all users have some preferences, defined as a set of keywords. Those preferences might or might not be visible to the OppNets application or algorithms. Furthermore, we allow each user to make a random decision about whether she likes the message, based on her preferences. After the simulation has finished, we let the simulated user evaluate each created message, irrespective of whether she actually received it or not, and if received, when. Later we can compare this list with the actual received messages and compute the user-based delivery rate. Furthermore, if some of the messages were received too late, we exclude them from the delivered ones. This idea can be extended to also consider popular, but late messages.

Another challenge to be addressed is the perceived quality of service (QoS) for a particular user. While the average delivery rate and delay might be very good for a particular scenario or network, there might be some users who never received any messages, or with much higher delays than the average. These outliers need to be identified and examined carefully as well.

% !TEX root = report-ieee.tex

\section{Conclusion}\label{sec:conc}

In this survey we have explored all relevant simulation models and tools for analyzing the performance of Opportunistic Networks. We have offered a taxonomy for most of the available models and we have explored their impact on the analysis results and their performance. We have compared the four most important OppNets simulators in terms of their performance, user friendliness and available models.

Furthermore, we have identified lacks in current models and have proposed several concrete improvements. We encourage the community to take these ideas, develop them further and thus contribute to more sophisticated, realistic and comparable research in Opportunistic Networks.

\section*{Acknowledgment}% IEEE prefers singular form
This work was partially supported by the {\em Ministerio de Economia y Competitividad, Programa Estatal de Investigaci\'{o}n, Desarrollo e Innovaci\'{o}n Orientada a los Retos de la Sociedad, Proyectos I+D+I 2014}, Spain, under Grant TEC2014-52690-R and by the Universidad Laica Eloy Alfaro de
Manab\'{i} (ULEAM), and the Secretar\'{i}a Nacional de Educaci\'{o}n Superior, Ciencia, Tecnolog\'{i}a e Innovaci\'{o}n (SENESCYT), Ecuador.

\bibliographystyle{IEEEtran}
\bibliography{misc}

\appendices
\onecolumn
\begin{landscape}
\section{Sample Experiment Journal}\label{app:log}
    \resizebox{\columnwidth}{!}{%
        \begin{tabular}{|m{12em}|m{9em}|m{9em}|m{9em}|m{9em}|m{9em}|m{9em}|m{9em}|m{5em}|m{4em}|m{5em}|m{8em}|}
        \hline
        \multicolumn{12}{|c|}{\cellcolor{black!5}\textbf{General}}\\
        \hline
        \textbf{Experimental goal}&\multicolumn{11}{l|}{Compare SWIM with SFO traces and RWP}\\
        \hline
        \textbf{Dates} (start - end)&\multicolumn{11}{l|}{Feb, 25th, 2017 - Feb 26th, 2017}\\
        \hline
        \textbf{Machine}&\multicolumn{11}{l|}{hebe virtual machine, Ubuntu 16.10, 4.8 standard kernel, 8 processor cores, 48 GB RAM}\\
        \hline
        \textbf{Simulation tool version}&\multicolumn{11}{l|}{OMNeT++ 5.0}\\
        \hline
        \textbf{Simulation tool add-ons, framework etc.\ version}&\multicolumn{11}{l|}{INET 3.4.0, OPS 0.1}\\
        \hline
        \textbf{Final logs archive location}&\multicolumn{11}{l|}{earth:/dev/null}\\
        \hline
        \textbf{Responsible}&\multicolumn{11}{l|}{Asanga}\\
        \hline
        \multicolumn{12}{|c|}{\cellcolor{black!5}\textbf{Simulation Models}}\\
        \hline
        \textbf{Application}&\multicolumn{11}{l|}{Personal message passing, people sending messages to each other}\\
        \hline
        \textbf{Mobility model}&\textbf{No.\ of nodes}&\textbf{Trace resolution}&\textbf{Position update resolution}&\textbf{Trace duration}&\textbf{Mean speed}&\textbf{Pause time}&\textbf{Environment size}&\textbf{Number of locations}&\textbf{alpha (SWIM)}&\textbf{Comments}&\textbf{Implementation}\\
        \hline
        \textit{SFO traces}&50&1 sec&1 sec&24 days&inherent&inherent&inherent&inherent&n/a&&BonnMotion\\
        \greyhline
        \textit{SWIM}&50&n/a&1 sec&24 days&12.2 m/s&300 s&45000~m $\times$ 45000~m&n/a&n/a&&self-implemented, published\\
        \greyhline
        \textit{Random Waypoint}&50&n/a&1 sec&24 days&12.2 m/s&300 s&45000~m $\times$ 45000~m&n/a&n/a&&INET\\
        \hline
        \textbf{Radio propagation model}&\textbf{Transmission Radius}&Path loss exponent&&&&&&&&&\textbf{Implementation}\\
        \hline
        \textit{UDG}&50 m&n/a&&&&&&&&&OPS 0.1\\
        \hline
        \textbf{Interference model}&\textbf{Interference Radius}&&&&&&&&&&\textbf{Implementation}\\
        \hline
        \textit{None}&n/a&&&&&&&&&&n/a\\
        \hline
        \textbf{Link technology}&\textbf{Scan Interval}&\textbf{Protocol}&\textbf{Bandwidth}&\textbf{Link delay}&\textbf{Cache size}&\textbf{Cache del.\ strategy}&\textbf{Cache add.\ strategy}&&&&\textbf{Implementation}\\
        \hline
        \textit{Simple direct-contact}&1 s&n/a&2.1 Mbps&0&infinite&n/a&LIFO&&&&OPS 0.1\\
        \hline
        \textbf{Data Propagation model}&\textbf{Beacon Interval}&\textbf{TTL}&\textbf{Cache size}&\textbf{cache del.\ strategy}&\textbf{cache add.\ strategy}&&&&&&\textbf{Implementation}\\
        \hline
        \textit{RRS}&1 s&48 h&infinite&oldest first&n/a&&&&&&OPS 0.1\\
        \hline
        \textbf{User behavior model}&\textbf{Change mobility}&\textbf{Reply to messages}&&&&&&&&&\textbf{Implementation}\\
        \hline
        \textit{None}&n/a&n/a&&&&&&&&&n/a\\
        \hline
        \textbf{Traffic model}&\textbf{Shape (const, uniform, posson etc.)}&\textbf{Mean number of messages per hour}&\textbf{Message size}&\textbf{Destination}&&&&&&&\textbf{Implementation}\\
        \hline
        \textit{Constant}&constant&2&1 KB&random single destination&&&&&&&OPS 0.1\\
        \hline
        \textbf{Energy consumption model}&\textbf{Energy per message}&\textbf{Energy per bit}&&&&&&&&&\textbf{Implementation}\\
        \hline
        \textit{None}&n/a&n/a&&&&&&&&&n/a\\
        \hline
        \textbf{Battery model}&\textbf{Capacity}&&&&&&&&&&\textbf{Implementation}\\
        \hline
        \textit{None}&n/a&&&&&&&&&&n/a\\
        \hline
        \multicolumn{12}{|c|}{\cellcolor{black!5}\textbf{Metrics}}\\
        \hline
        &\textbf{Network-wide}&\textbf{Mean per node}&\textbf{Full logs per node / message}&&&&&&&&\\
        \hline
        \textit{Delivery rate for all messages}&-&-&&&&&&&&&\\
        \greyhline
        \textit{Delivery rate for wished messages only}&computed&computed&\checkmark&&&&&&&&\\
        \greyhline
        \textit{Delivery delay for all messages}&-&-&&&&&&&&&\\
        \greyhline
        \textit{Delivery delay for wished messages only}&computed&computed&\checkmark&&&&&&&&\\
        \greyhline
        \textit{Individual contacts}&computed&computed&\checkmark&&&&&&&&\\
        \greyhline
        \textit{Messages received (as destination)}&computed&computed&\checkmark&&&&&&&&\\
        \greyhline
        \textit{Messages created}&computed&computed&\checkmark&&&&&&&&\\
        \greyhline
        \textit{Messages forwarded}&computed&computed&\checkmark&&&&&&&&\\
        \greyhline
        \textit{Messages dropped}&n/a&n/a&n/a&&&&&&&&\\
        \greyhline
        \textit{Energy spent}&-&-&-&&&&&&&&\\
        \greyhline
        \textit{Simulation duration (wall clock time)}&\checkmark&-&-&&&&&&&&\\
        \greyhline
        \textit{Simulation memory}&\checkmark&-&-&&&&&&&&\\
        \hline
        \multicolumn{12}{|c|}{\cellcolor{black!5}\textbf{Comments and Notes}}\\
        \hline
        \multicolumn{12}{|l|}{Trace files were created in real time on file server (earth)}\\
        \multicolumn{12}{|l|}{Wall clock time measured using \textit{time} command}\\
        \multicolumn{12}{|l|}{RAM usage measured using \textit{pidstat -r -u -C}}\\
        \hline
    \end{tabular}
}
\end{landscape}
\twocolumn

\end{document}